%% file: High-z_UV_absorption.tex
\documentclass[twocolumn]{aastex701}
\let\tablenum\relax

\usepackage{url}
\usepackage{svg}
\usepackage{threeparttable}
\usepackage{ulem}
\usepackage{natbib}
\usepackage{appendix}
\usepackage{amsmath}
\usepackage{amssymb}
\usepackage{multirow}
\usepackage{float}
\usepackage{physics}
\usepackage{siunitx}
\usepackage{mhchem}
\usepackage{appendix}

\def\ion#1#2{#1$\;${\sc\@roman{#2}}\relax}
\def\lesssim{\mathrel{\hbox{\rlap{\hbox{\lower4pt\hbox{$\sim$}}}\hbox{$<$}}}}
\def\gtrsim{\mathrel{\hbox{\rlap{\hbox{\lower4pt\hbox{$\sim$}}}\hbox{$>$}}}}

\def\red{\textcolor{black}}

\shorttitle{Absorption Lines and Chemical Enrichment of Galaxies at $z=7.2-10.6$}
\shortauthors{Nakane et al.}
\begin{document}
\title{
JWST Absorption-Line Analysis of UV-Bright Galaxies at $z=7.2-10.6$:\\
Early Chemical Enrichment Traced by C, O, Mg, Al, Si, and Fe
%
%Chemical Evolution Probed with C, O, Mg, Al, Si, and Fe
%
%Emergence of Fe-Enriched Cool Gas in UV-bright Galaxies at $z=7.2-10.6$
}

\input{author_list}



\begin{abstract}
%230 words
%We investigate UV absorption lines tracing cool gas in eight bright ($-21.9<M_\mathrm{UV}<-20.5$) galaxies at $z=7.2-10.6$ using deep JWST/NIRSpec medium-resolution spectra homogeneously selected from archival data.
% \textcolor{red}{
We investigate UV absorption lines tracing cool gas in eight bright ($-21.9<M_\mathrm{UV}<-20.5$) galaxies at $z=7.2-10.6$ using deep \red{archival JWST/NIRSpec medium-resolution spectra. The eight galaxies were homogeneously selected based on wavelength coverage and data quality from $362$ galaxies at $z>7$ available to date, and all eight were observed as part of the SPURS program.} We identify multiple absorption lines of low-ionization species (LIS) in five galaxies, while the remaining three show only one or no significant detections. The average LIS equivalent widths $\mathrm{EW_{LIS}}$ indicate weaker absorption lines in compact ($r_\mathrm{e}\lesssim 100$ pc) galaxies with high $L_\mathrm{H\beta}/L_\mathrm{UV}$ and $\Sigma_\mathrm{SFR}$, suggestive of absorption-line suppression due to intense star formation or nuclear activity. We simultaneously fit multiple LIS lines, accounting for both covering fraction $C_f$ and column density $N$, and find that the five galaxies follow the local $C_f$-$N$ relation. Through the LIS-line fitting, we derive chemical abundance ratios. The five galaxies have abundance ratios similar to those of damped Ly$\alpha$ (DLA) systems \red{at $z\sim5-7$} in the [Si/O]-[C/O] and [Fe/Si]-[Al/Si] planes, suggesting that their absorbing gas is similar to that of DLAs. We find that two galaxies at $z\sim 7$ are already highly Fe-enriched, with $\mathrm{[Fe/Si]}\simeq -0.2$, near the solar abundance ratio. While the high Fe enrichment may originate from early Type-Ia supernovae (SNe Ia) or pair-instability supernovae (PISNe), these two galaxies also show $\mathrm{[Al/Si]}\sim0$, with no signature of the strong odd-even effect expected from PISNe, \red{in which odd-$Z$ elements are suppressed relative to even-$Z$ elements,} favoring SNe Ia. 
\end{abstract}

\keywords{Chemical enrichment (225); Early universe (435); Galaxy chemical evolution (580); Galaxy evolution (594); Galaxy formation (595); High-redshift galaxies (734); Interstellar line absorption (843)}

%% subsec tion for cross-referencing in a \ref command.

\section{Introduction}
\label{sec:introduction}
James Webb Space Telescope (JWST) has identified many high-redshift galaxies up to $z>14$ (e.g., \citealt{Bunker2023,Curtis-Lake2023,Carniani2024,Castellano2024,Naidu2025,Donnan2026}), revealing the unusual nature of UV-bright galaxies at $z\gtrsim6$, such as extremely compact morphologies, high-ionization emission lines, and extreme nitrogen enhancement (e.g., \citealt{Cameron2023,Isobe2023b,Castellano2024,Maiolino2024,Senchyna2024,Topping2024,Topping2025a}). In contrast, some UV-bright galaxies with extended morphologies show weak emission lines (e.g., \citealt{Hainline2024,Harikane2025,Chen2026a,Donnan2026,Harikane2026}), demonstrating variations in their properties. The detections of UV-to-optical continuum and emission lines have enabled us to explore the stellar populations and hot H \textsc{ii} gas ionized by young massive stars for these UV-bright galaxies. On the other hand, UV low-ionization species (LIS) absorption lines trace cool neutral gas associated with the interstellar medium (ISM) and circumgalactic medium (CGM; e.g., \citealt{Shapley2003,Jones2013}). This cool gas provides unique insights into feedback-driven outflows (e.g., \citealt{Steidel2010}) and chemical enrichment histories that may differ from stars or ionized gas (e.g., \citealt{Jones2018,Christensen2023,Sodini2024}). Despite the rapid progress in characterizing stellar populations and ionized gas properties with JWST, our understanding of cool gas properties in galaxies during the epoch of reionization remains limited.

In previous studies of UV absorption lines in lower-redshift galaxies, the strengths and velocity profiles of LIS lines have been used to constrain gas covering fractions and feedback-driven outflows (e.g., \citealt{Shapley2003,Steidel2010,Jones2013,Shibuya2014,Sugahara2019,VasanGC2025}). In addition, weak LIS absorption is associated with strong Ly$\alpha$ emission, suggesting a potential connection with leakage of ionizing photons (e.g., \citealt{Jones2013,Chisholm2018}). Studies of galaxy samples at $z\sim2-5$ suggest a possible redshift evolution of UV absorption lines (weaker absorption at higher redshifts; e.g., \citealt{Du2018,Pahl2020}). Beyond gas kinematics and covering fractions, UV absorption lines have been used to constrain elemental abundances (e.g., C, O, Si, and Fe) in the cool gas of $z\sim2-3$ star-forming galaxies \citep{Jones2018} and damped Ly$\alpha$ (DLA) systems at $z\sim5-7$ along the sight lines of quasars (QSOs; \citealt{Sodini2024}). However, UV absorption line studies were limited to the low-redshift galaxies or QSO absorption systems due to the requirement for UV continuum detections with high signal-to-noise (S/N) ratios. The high sensitivity of JWST/NIRSpec \citep{Jakobsen2022} has enabled observations of UV absorption lines even during the epoch of reionization ($z\gtrsim6$). Recent studies also show weaker LIS absorption in high-z galaxies compared to lower-z galaxies \citep{Glazer2025,Chen2026a,Pollock2026,VasanGC2026}. However, existing studies are primarily based on stacked spectra \citep{Glazer2025} or a limited number of individual galaxies \citep{Boyett2024,Topping2025a,Chen2026a,Pollock2026,VasanGC2026,Zhu2026}. In addition, despite the potential of UV absorption lines to constrain chemical abundance ratios of cool gas, their constraints remain limited especially at $z\gtrsim6$ primarily due to the difficulty of disentangling the effects of covering fraction and column density on the absorption-line profiles. This motivates the investigation of UV absorption lines in individual high-$z$ galaxies to characterize both the observed weakness of LIS absorption and the chemical enrichment of cool gas during the epoch of reionization.

In this work, we analyze UV absorption lines of eight individual galaxies at $z=7.2-10.6$, using deep NIRSpec medium-resolution spectra. This paper is constructed as follows. Section \ref{sec:sample_data} presents our sample galaxies and spectroscopic/photometric data used in this study. In Section \ref{sec:analysis}, we describe the method of our analysis in detail. We provide our results in Section \ref{sec:result}. In Section \ref{sec:discussion}, we discuss origins of weak LIS absorption line and iron enrichment of high-$z$ galaxies. Section \ref{sec:summary} summarizes our major findings. We assume a standard $\Lambda$CDM cosmology with $\Omega_\Lambda=0.7$,$\Omega_m=0.3$, and $H_0=70\ \mathrm{km}\ \mathrm{s^{-1}}\ \mathrm{Mpc^{-1}}$. All magnitudes are in the AB system \citep{Oke&Gunn1983}. Throughout this paper, we utilize the solar abundance ratios of \citet{Asplund2021}. The notation [X/Y] is defined as $\log(X/Y)$ relative to the solar abundance ratio, i.e., $[X/Y]=\log(X/Y)-\log(X/Y)_\odot$.

\section{Sample and Data}
\label{sec:sample_data}

\subsection{Sample}
\label{subsec:sample}
To detect UV absorption lines, we require \red{sufficient continuum sensitivity and spectral resolution to separate neighboring transitions.}  We thus construct our sample using JWST/NIRSpec medium-resolution ($R\sim1000$) grating spectra. We use the publicly available data from the DAWN JWST Archive (DJA ver 4.4 \footnote[1]{\url{https://dawn-cph.github.io/dja/spectroscopy/nirspec/}}; \citealt{Heintz2025,de_Graaff2025,Valentino2025}). We also utilize the data obtained in JWST Cycle 4 program, SPectroscopic Ultra-deep Reionization-era Survey (SPURS; GO-9214; PIs: C. Mason \& D. Stark; \citealt{Chen2026a}), motivated by recent reports of UV absorption line detections in individual galaxies at $z>7$ from the SPURS data \citep{Chen2026a,Zhu2026}. \red{We first compile a total of $362$ galaxies at $z>7$ with medium-resolution spectra from the DJA (grade 3) and SPURS datasets. We then select $102$ galaxies observed with grating/filter pairs of G140M/F070LP or G140M/F100LP, which cover key rest-frame UV absorption lines, such as Si \textsc{ii} $\lambda1260$ and O \textsc{i} $\lambda1302$. We adopt $z>7$ to maximize the number of UV absorption transitions available for robust column-density measurements.} We calculate the median continuum S/N ratios per $10$ \AA\ in the observed frame over the rest-frame wavelength range of $1250-1700$ \AA, masking any emission and absorption features.
\begin{figure}[ht!]
    \centering
    \includegraphics[width=0.9\linewidth]{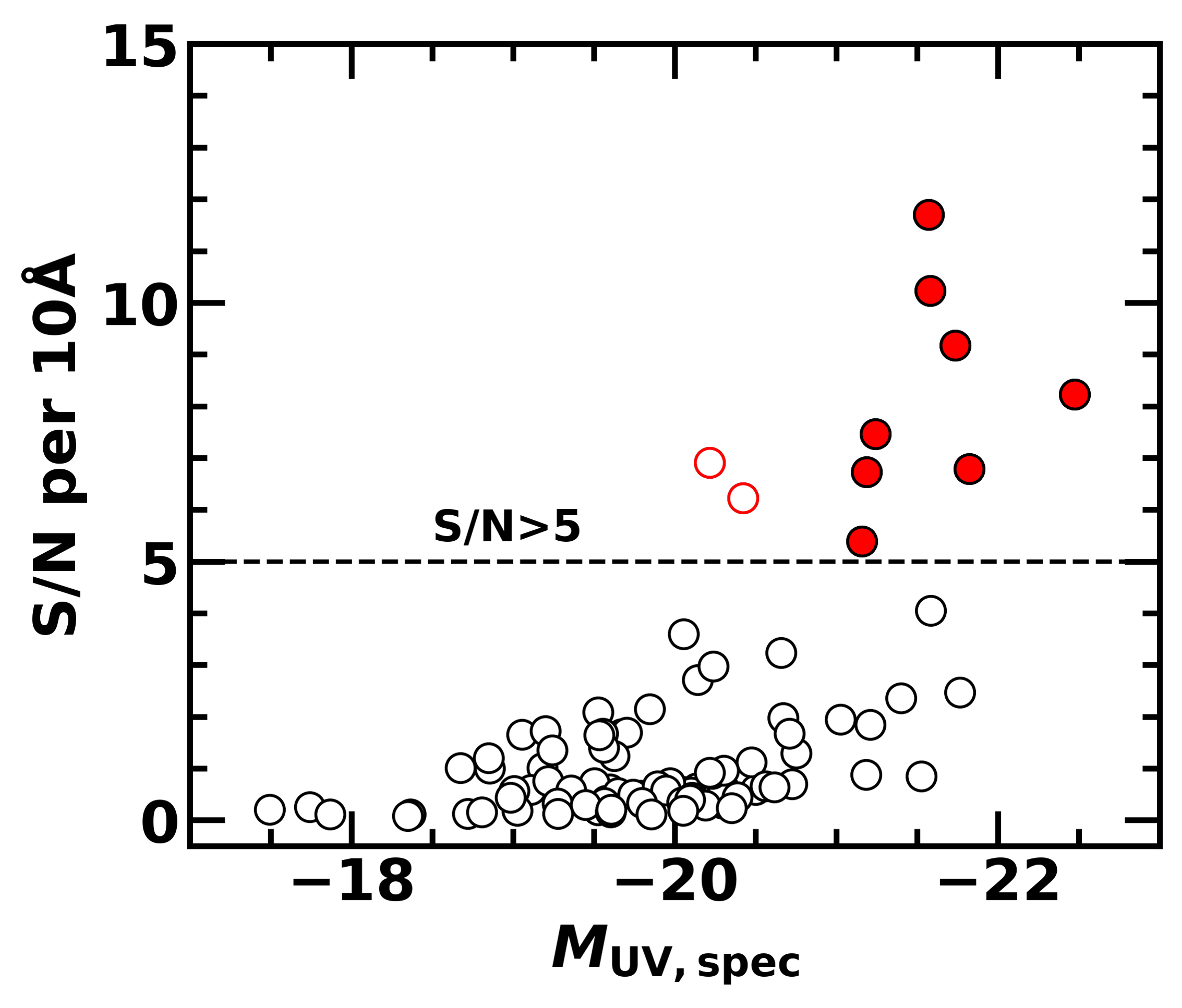}
    \caption{Signal-to-noise ratio \red{of the UV continuum} per $10$ \AA\ in the observed frame as a function of spectroscopic UV magnitude for galaxies used in our sample selection. The S/N and $M_\mathrm{UV,spec}$ values are derived over the rest-frame wavelength range of $1250-1700$ \AA. The red filled circles show our sample galaxies with $\mathrm{S/N}>5$. The red open circles denote galaxies with $\mathrm{S/N}>5$ excluded from our sample due to the significant detection of \red{broad Balmer emission indicative of an AGN} or lack of wavelength coverage for key absorption lines. The black open circles present galaxies with \red{$\mathrm{S/N}<5$}.}
    \label{fig:M_UV-SN}
\end{figure}
In Figure \ref{fig:M_UV-SN}, we present the continuum S/N ratios of the $102$ galaxies with spectroscopic UV magnitudes. We set the following criteria: $\mathrm{S/N}>5$ per $10$ \AA\ in the observed frame, sufficient wavelength coverage of the key absorption lines, and \red{no significant broad Balmer emission indicative of a broad-line active galactic nucleus (AGN).} We select a final sample of $8$ galaxies (GN-z11, Gz9p3, CEERS-1025, CEERS-1019, GLASS-100003, GLASS-10021, GN-z7p2, and EGS-z7p2) at $z=7.19-10.60$ with UV magnitudes of $-21.9<M_\mathrm{UV}<-20.5$ (see Section \ref{sec:analysis} for the redshift and UV magnitude measurement), as shown in Figure \ref{fig:z-M_UV}. Our sample galaxies are among the UV-brightest galaxies known at these redshifts. We summarize our sample in Table \ref{tab:sample}. We note that our sample includes some galaxies (Gz9p3, CEERS-1019, GLASS-100003, and GLASS-10021), for which UV absorption lines have been reported and analyzed in previous studies \citep{Boyett2024,Chen2026a,Pollock2026,VasanGC2026,Zhu2026}.

\begin{figure}[ht!]
    \centering
    \includegraphics[width=0.99\linewidth]{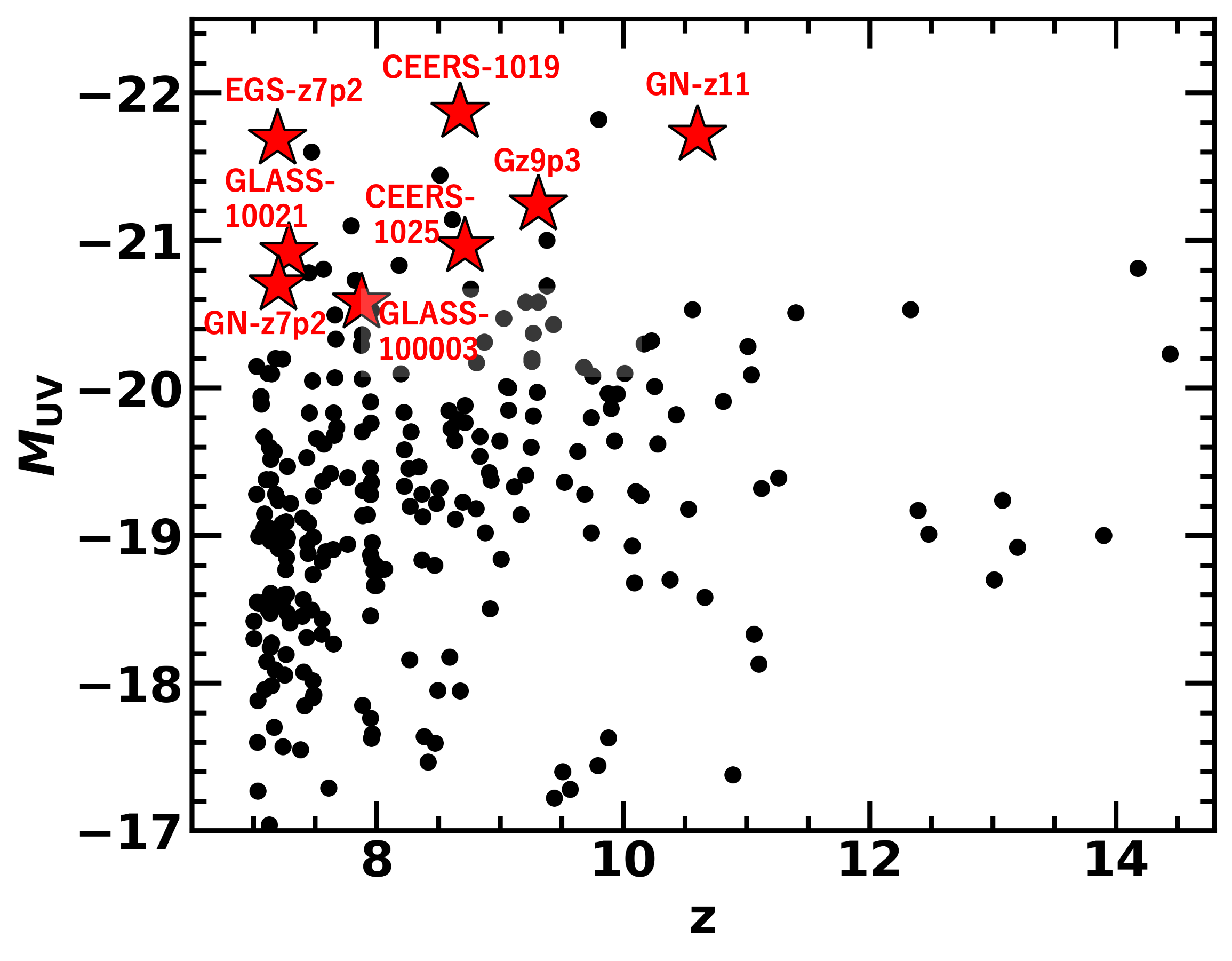}
    \caption{Absolute UV magnitude distribution as a function of redshift. The red star symbols denote our sample galaxies while the black circles indicate other spectroscopically confirmed galaxies at $z>7$ compiled from literature \citep{Haro2023a,Haro2023b,Curtis-Lake2023,Nakajima2023,Bunker2024,Carniani2024,Castellano2024,D'Eugenio2024,Fujimoto2024,Harikane2024,Roberts-Borsani2024,Harikane2025,Kokorev2025,Naidu2025,Napolitano2025,Tang2025,Witstok2025}. The UV magnitudes are corrected for magnification.}
    \label{fig:z-M_UV}
\end{figure}
\input{Table/table_sample}

\begin{figure*}[ht!]
    \centering
    \includegraphics[width=0.99\linewidth]{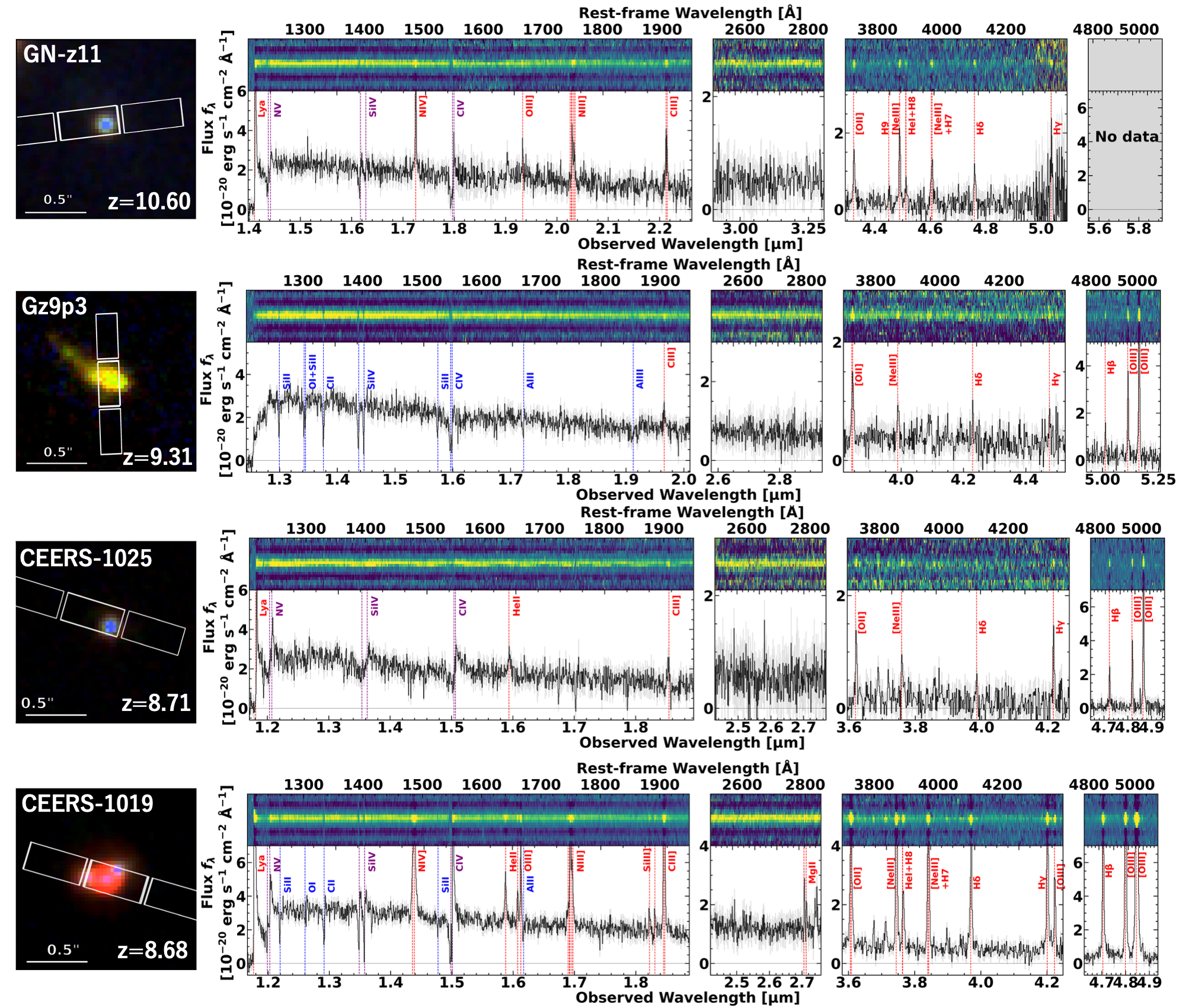}
    \caption{NIRSpec spectra (right) and NIRCam $1\farcs5\times1\farcs5$ color images (left; R: F444W, G: F277W, B: F150W) of our sample galaxies. In the left panels, the white rectangles indicate the MSA slit positions. The right panels present the two-dimensional spectra (top) and one-dimensional spectra (bottom). The red and blue dashed lines present detected emission and absorption lines, respectively. The purple dashed lines denote the \red{stellar wind signatures}, including those reported by \citet{Chen2026b,Marques-Chaves2026,Nakane2026}.}
    \label{fig:sample}
\end{figure*}
\setcounter{figure}{2}
\begin{figure*}[ht!]
    \centering
    \includegraphics[width=0.99\linewidth]{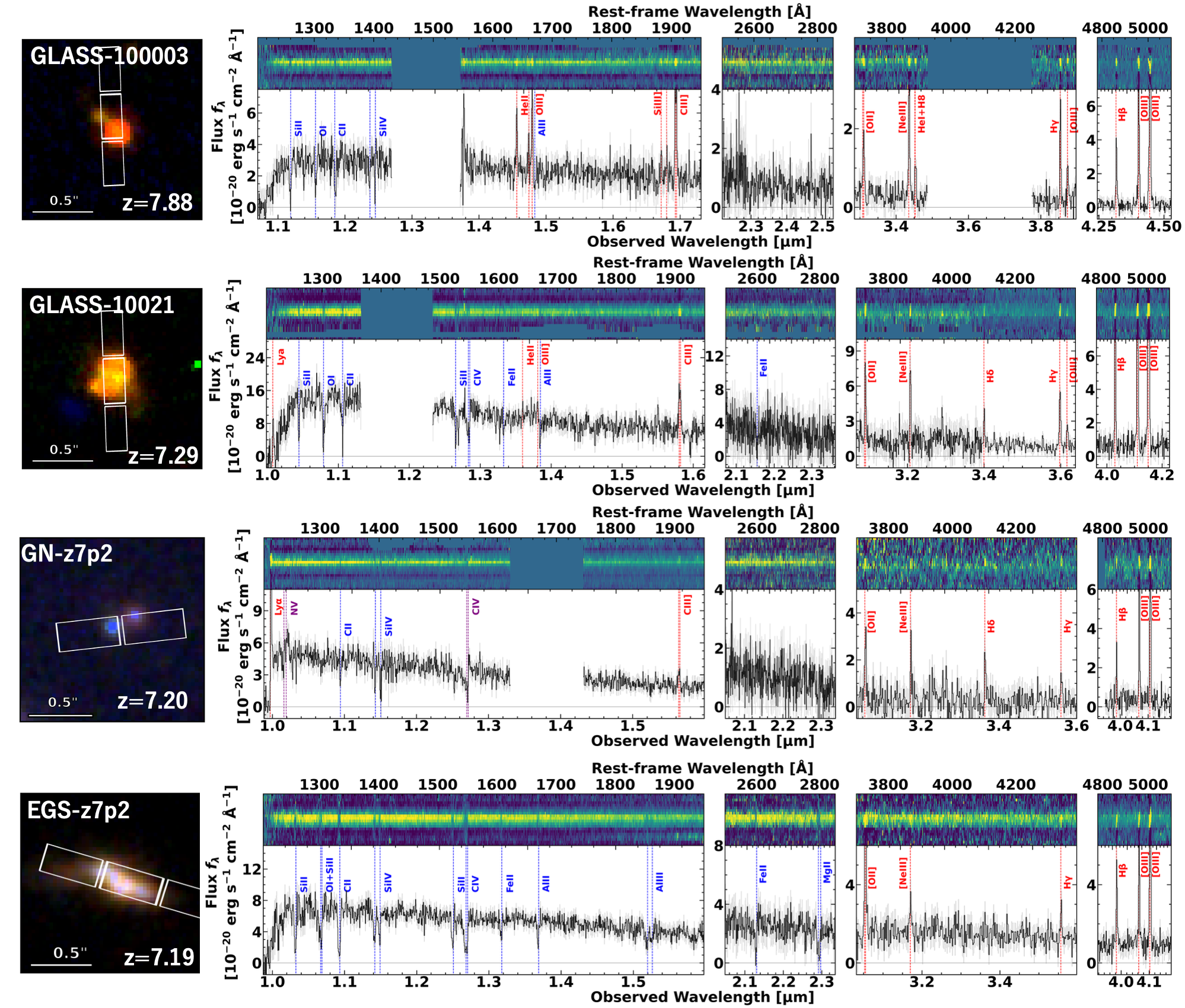}
    \caption{Continued.}
    \label{fig:sample}
\end{figure*}

\subsection{Spectroscopic Data}
\label{subsec:nirspec}
While candidate galaxies are initially identified from both the DJA and SPURS datasets, all galaxies that satisfy our selection criteria have deep NIRSpec observations from SPURS. Therefore, we base our spectroscopic analysis solely on the SPURS data. The SPURS data were taken with JWST/NIRSpec Micro-Shutter Assembly (MSA; \citealt{Ferruit2022,Jakobsen2022}) in the Abell 2744 cluster (6-9 November 2025), EGS (29 February - 1 March 2026), and GOODS-N (12-14 May 2026) fields. The NIRSpec observations were conducted using the medium-resolution grating/filter pairs of G140M/F100LP, G235M/F170LP, and G395M/F290LP, covering $1.0-1.6$, $1.7-3.1$, and $2.9-5.1$ $\mu$m, respectively. The total exposure times for the G140M, G235M, and G395M gratings are $29.2$, $7.9$, and $2.9$ hours, respectively. The spectra are reduced with \texttt{msaexp} (v0.9.18; \citealt{Brammer2023}) based on the JWST pipeline (ver.1.16.0\red{; \citealt{Bushouse2024}}) with the Calibration Reference Data System (CRDS) \texttt{jwst\_1322.pmap} calibration file, following the procedures in previous works \citep{de_Graaff2024,Heintz2025,Valentino2025}.

\subsection{Photometric Data}
\label{subsec:nircam}
We use the HST and JWST/NIRCam imaging data, point spread functions (PSFs), and photometric catalogs of our sample galaxies provided by Ultra-deep NIRCam and NIRSpec Observations Before the Epoch of Reionization (UNCOVER; GO-2561, PI: I. Labbe \& R. Bezanson; \citealt{Bezanson2024}), Medium Bands, Mega Science (MegaScience; GO-4111, PI: K. Suess; \citealt{Suess2024}), Cosmic Evolution Early Release Science (CEERS; ERS-1345, PI: S. Finkelstein; \citealt{Haro2023a,Finkelstein2023,Cox2025}), and JWST Advanced Deep Extragalactic Survey Data Release 5 (JADES DR5 \citealt{Eisenstein2026,Johnson2026,Robertson2026}; GTO-1180, 1181, 1210, 1264, 1286, 1287, 4540; GO-1895, PI: P. Oesch \citealt{Oesch2023} 2514, PI: C. Williams, \citealt{Williams2025} 2674, PI: P. Arrabal Haro, 3577, PI: E. Egami, 4762, PI: S. Fujimoto, 5398, PI: J. Kartaltepe, and 6434, PI: E. Egami, \citealt{Sun2025}). \red{The NIRCam/F200W imaging data and corresponding PSFs are used for the size measurements described in Section \ref{subsubsec:morphology}, while the published multi-band catalog photometry is used for the SED fitting in Section \ref{subsubsec:sed}. The catalog photometry accounts for wavelength-dependent PSF variations through PSF matching and/or aperture corrections as described in the respective survey papers.}

\section{Analysis}
\label{sec:analysis}
\subsection{Spectroscopic Analysis}
\label{subsec:spec}
\subsubsection{Systemic Redshift Determination}
\label{subsubsec:z_sys}
We determine systemic redshifts ($z_\mathrm{sys}$) of our sample \red{galaxies} by fitting to the [O \textsc{iii}] $\lambda\lambda4959,5007$ doublet lines. For GN-z11, whose [O \textsc{iii}] emission is not covered by the NIRSpec wavelength range, we adopt the literature value based on multiple nebular emission lines observed with NIRSpec ($z_\mathrm{sys}=10.6034$; \citealt{Bunker2023}). We first conduct single-component fitting with two Gaussian functions for emission lines, for which we fix the center wavelength separation based on the rest-frame wavelengths and flux ratio to be $F_\mathrm{[OIII]5007}/F_\mathrm{[OIII]4959}=2.98$ \citep{Storey&Zeippen2000}, and a linear function for a continuum to determine a local continuum level. To account for the instrumental broadening, we assume the relation $\mathrm{FWHM}_\mathrm{int}=\sqrt{\mathrm{FWHM}_\mathrm{obs}^2-\mathrm{FWHM}_\mathrm{inst}^2}$, where $\mathrm{FWHM}_\mathrm{int}$, $\mathrm{FWHM}_\mathrm{obs}$, and $\mathrm{FWHM}_\mathrm{inst}$ are intrinsic, observed, and instrumental full width at half maximum (FWHM), respectively. To derive $\mathrm{FWHM}_\mathrm{inst}$, we approximate the line spread functions (LSFs) derived by \citet{Isobe2023a} as Gaussian profiles. Because the effective NIRSpec resolution for compact sources is approximately twice the nominal resolution, we scale the LSF widths by a factor of $0.5$. Some of our sample galaxies show outflow features, including broad \red{($\mathrm{FWHM}\sim650$ km s$^{-1}$)} outflow components of [O \textsc{iii}] $\lambda\lambda4959,5007$ and H$\beta$ in CEERS-1019 \citep{Zamora2025} and blue-shifted UV absorption lines in Gz9p3, GLASS-100003, and GLASS-10021 \citep{Chen2026a,VasanGC2026,Zhu2026}. Therefore, we add a second Gaussian component representing a possible outflow feature, for which we fix the center wavelength separation and flux ratio. To obtain posterior distributions of the fitting parameters, we conduct Markov Chain Monte Carlo (MCMC) simulations using \texttt{emcee} \citep{Foreman2013}. We apply flat priors to the free parameters of the line flux $F_\mathrm{[OIII]5007}$, line center wavelength $\lambda_\mathrm{cen,[OIII]5007}$, intrinsic line width $\mathrm{FWHM}_\mathrm{int}$, slope $a_\mathrm{cont}$, and intercept $b_\mathrm{cont}$ of the linear function for the single-component fitting. For the double-component fitting, we add three parameters of the second Gaussian components with flat priors. We evaluate the goodness of the single- and double-component fits using the widely applicable information criterion (WAIC; \citealt{Watanabe2010}), which is applicable to Bayesian parameter estimation. We adopt double-component fitting results with the two criteria: 1) $\Delta\mathrm{WAIC}=\mathrm{WAIC\ (double\ components)}-\mathrm{WAIC\ (single\ component)}<-11.8$ (corresponding approximately to $3\sigma$ significance level; e.g., \citealt{Hviding2025})  and 2) $\mathrm{S/N}>5$ for the outflow components. Based on these criteria, we detect outflow components for CEERS-1019 with $\Delta$WAIC=$-481.4$ and $\mathrm{S/N}=13.8$, which is consistent with the previous results \citep{Zamora2025}. While we do not identify statistically significant outflow components for other sample galaxies, we note that Gz9p3, CEERS-1025, GLASS-100003, and GLASS-10021 show tentative outflow signatures. We determine the best-fit parameters and their $1\sigma$ uncertainties from the mode (i.e., the peak of the posterior distribution) and the $68\%$ highest posterior density interval (HPDI; i.e., the narrowest interval containing $68\%$) of the preferred fit based on the WAIC values. We derive $z_\mathrm{sys}$ from the center wavelength of the main component. 
%In Table XX, we summarize the fitting results.

\subsubsection{Line Flux Measurement}
\label{subsubsec:line_flux}
We measure emission and absorption line fluxes by fitting Gaussian functions and a linear function in a similar manner to Section \ref{subsubsec:z_sys}. To reduce degeneracies in the flux measurements of weak lines, we adopt a two-step fitting procedure.  We first conduct the fitting with all parameters free. For weak lines with $\mathrm{S/N}<4$, we repeat the fit with the velocity offset and line width fixed to those of the main [O \textsc{iii}] $\lambda\lambda4959,5007$ component, or the average values of other securely detected absorption lines, where available. Lines with $\mathrm{S/N}>3$ are considered detections, while the remaining lines are treated as non-detections. For multiple transitions from the same ion at nearby wavelengths (e.g., Si \textsc{iv} $\lambda\lambda1393,1402$; C \textsc{iii}] $\lambda\lambda1907,1909$), we use a common velocity offset and FWHM in velocity units. For Ly$\alpha$ line, we account for the intergalactic medium (IGM) absorption, multiplying the continuum and emission line models by the Ly$\alpha$ transmission models \citep{Inoue2014}. Although we detect outflow features of [O \textsc{iii}] emission lines in CEERS-1019 (see Section \ref{subsubsec:z_sys}), we do not conduct double-component fitting to other emission lines except for the H$\beta$ and Ly$\alpha$ lines, which also show significant broad components. We derive a rest-frame equivalent width (EW) of absorption lines with 
\begin{equation}
    \mathrm{EW}=\frac{F_\mathrm{line}}{f_\mathrm{cont}(1+z_\mathrm{sys})},
    \label{eq:EW}
\end{equation}
where $F_\mathrm{line}$ and $f_\mathrm{cont}$ are line flux and continuum flux density at the line center wavelength in the observed frame, respectively. We perform MCMC fitting using \texttt{emcee} and flat priors for the flux $F_\mathrm{line}$ (equivalent width $\mathrm{EW_{line}}$), line center wavelength $\lambda_\mathrm{cen}$, $\mathrm{FWHM_{int}}$, $a_\mathrm{cont}$, and $b_\mathrm{cont}$ for emission (absorption) lines. As summarized in Table \ref{tab:line}, we detect multiple emission and absorption lines with $\mathrm{S/N}>3$ in our sample galaxies. For non-detected lines with $\mathrm{S/N}<3$, we derive $3\sigma$ upper (lower) limits on the line fluxes (EWs) from the posterior distributions.

\subsubsection{Absorption Modeling}
\label{subsubsec:abs_model}

\begin{figure*}[ht!]
    \centering
    \includegraphics[width=0.99\linewidth]{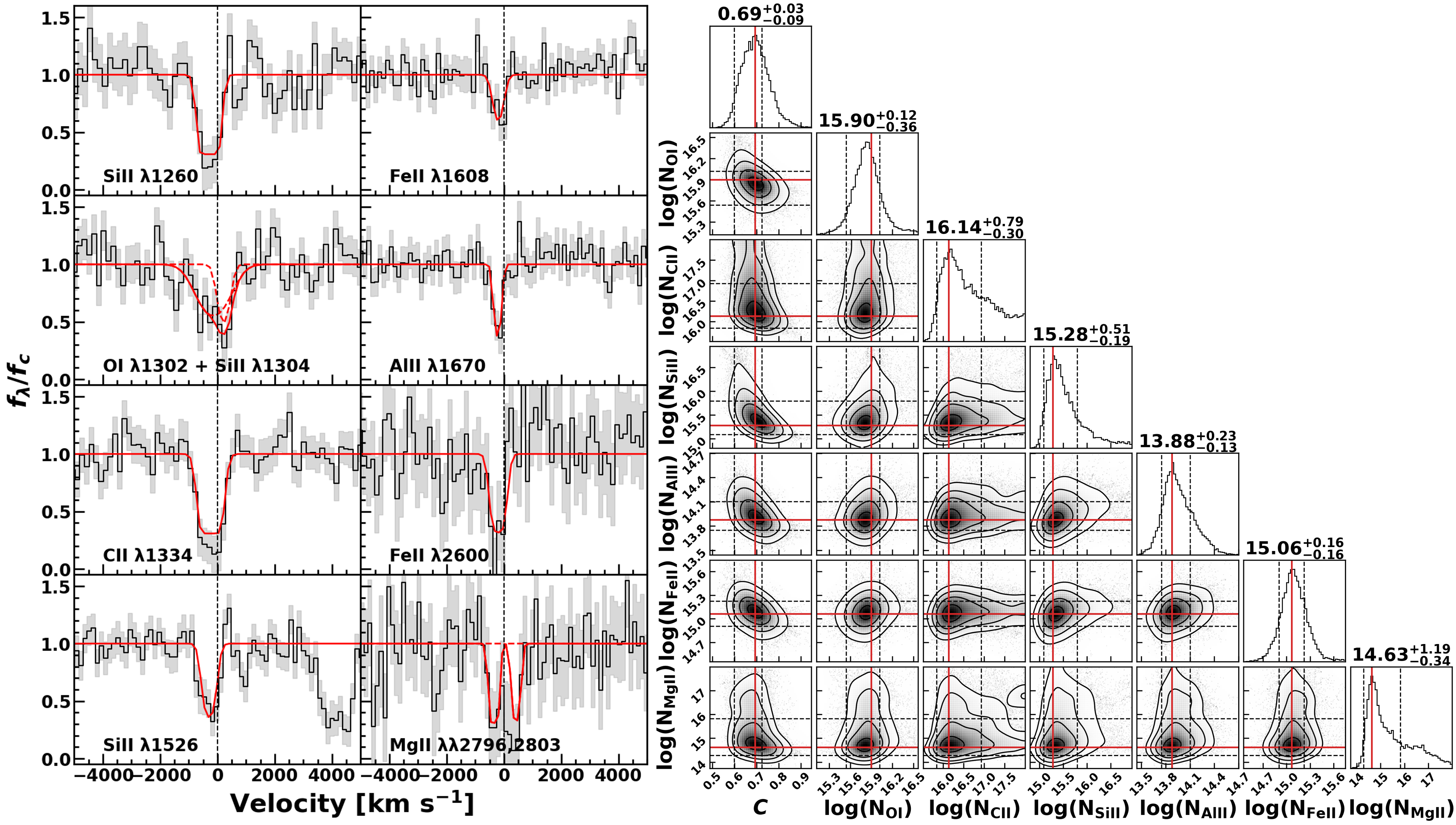}
    \caption{Example of PCM fitting results for EGS-z7p2. Left: the black lines, gray shaded regions, and red lines show the observed spectra, their $1\sigma$ uncertainties, and best-fit model spectra, respectively. Right: marginalized posterior distributions for covering fraction and ionic column densities. The red-solid and black-dashed lines represent the mode and boundaries of $68$th percentile HPDI, respectively.}
    \label{fig:PCM}
\end{figure*}

In Section \ref{subsubsec:line_flux}, we detect multiple UV absorption lines of LIS (Si \textsc{ii} $\lambda1260$, O \textsc{i} $\lambda1302$, Si \textsc{ii} $\lambda1304$, C \textsc{ii} $\lambda1334$, Si \textsc{ii} $\lambda1526$, Fe \textsc{ii} $\lambda1608$, Al \textsc{ii} $\lambda1670$, Fe \textsc{ii} $\lambda2600$, and Mg \textsc{ii} $\lambda\lambda2796,2803$) and high-ionization species (HIS; Si \textsc{iv} $\lambda\lambda1393,1402$, C \textsc{iv} $\lambda\lambda1548,1550$, and Al \textsc{iii} $\lambda\lambda1854,1862$). The LIS absorption lines trace cool and mostly neutral gas in the ISM/CGM in galaxies. On the other hand, the HIS absorption lines trace hot ionized gas that can arise from the ISM/CGM and galactic outflows, and may also include contributions from stellar winds driven by massive stars such as O-type and Wolf-Rayet stars (e.g., \citealt{Leitherer2011}). Due to their relatively simple ionization structure in neutral gas, LIS absorption lines have been used to measure abundance ratios of cool gas in the ISM/CGM (e.g., \citealt{Pettini2002}) and in absorption systems along QSO sight lines (e.g., \citealt{Lehner2016,Christensen2023,Sodini2024}). 

The abundance ratios in absorbing gas can be derived from LIS ionic column densities that is sensitive to absorption line depth. However, the absorption lines from optically thick gas are saturated and their depths are mostly determined by the covering fraction of the absorbing gas. Even for optically thin lines, the absorption line depth depends on both column density and covering fraction, making it difficult to uniquely constrain the column density from a single transition. To address the saturation and degeneracy between the column density and covering fraction, we use a standard partial covering model (PCM; e.g., \citealt{Jones2013}). The PCM assumes that the UV continuum from a galaxy is absorbed by partially covered absorbing gas as described with:
\begin{align}
    \frac{f_\lambda}{f_\mathrm{cont}}&=1-C_f+C_f e^{-\tau(\lambda)},
    \label{eq:PCM}
\end{align}
where $f_\lambda$, $f_\mathrm{cont}$, $C_f$, and $\tau(\lambda)$ represent the observed flux density, continuum flux density, covering fraction, and optical depth as a function of wavelength. We adopt a Gaussian profile for optical depth as:
\begin{align}
    \tau(\lambda)&=\tau_0 \exp\left(-\frac{(v(\lambda)-v_0)^2}{b_D^2}\right),\\
    \tau_0&=\frac{\pi e^2}{m_e c}\frac{\lambda_0 fN}{b_D},
    \label{eq:tau}
\end{align}
where $\tau_0$, $v(\lambda)$, $v_0$, $b_D$, $e$, $m_e$, $c$, $\lambda_0$, $f$, and $N$ are the optical depth at line center, velocity as a function of wavelength, velocity at line center, Doppler width, elementary charge, mass of electron, speed of light, rest-frame wavelength of lines, oscillator strength, and ionic column density. We adopt the oscillator strength values presented in \citet{Cashman2017}.

\input{Table/table_line}
\input{Table/table_absorption}

\noindent To simultaneously constrain column density and covering fraction, we fit the PCM for five galaxies (Gz9p3, CEERS-1019, GLASS-100003, GLASS-10021, and EGS-z7p2) with multiple LIS absorption lines detected. We include weak Si \textsc{ii} $\lambda1304$ line even if its S/N is low \red{(i.e., $\mathrm{S/N}<3$)}, because the detections of Si\textsc{ii} $\lambda\lambda1260,1526$ provide additional information on Si \textsc{ii} absorption, and Si \textsc{ii} $\lambda1304$ is partially blended with OI $\lambda1302$. If O \textsc{iii}] $\lambda\lambda1661,1666$ lines nearby Al \textsc{ii} $\lambda1670$ are detected, to avoid the contamination, we also include them to the PCM fitting, fixing the redshifts and line widths to those from [O \textsc{iii}] $\lambda\lambda4959,5007$. We fit the observed spectra in the rest-frame wavelength ranges of $1250-2850$ \AA, masking any other emission and absorption lines. Note that we exclude the observed wavelength range of $1.45-1.80\ \mu$m for EGS-z7p2, which is partially contaminated by the light from a nearby object. For the UV continuum, we use the stellar and nebular continuum models in \citet{Nakane2024b,Nakane2025}, which are constructed from the \red{stellar} population synthesis code \texttt{BPASS} v2.2.1 \citep{Eldridge2017,Stanway2018} and photoionization code \texttt{Cloudy} v23.01 \citep{Ferland1998,Gunasekera2023}. We adopt \red{a} \citet{Salpeter1955} initial mass function (IMF) with a high-mass cutoff of $100\ M_\odot$ including binary stars, and a constant star formation history, varying stellar metallicities of $Z_*=10^{-3}-0.040$ and stellar ages of $\log(t/\mathrm{yr})=6.0-11.0$. We use a \citet{Calzetti2000} extinction law parametrized by a color excess of $E(B-V)$, and the IGM absorption models of \citet{Inoue2014}. The model spectra are smoothed to match the observed spectral resolution of $R=1000$ for the grating spectra. See \citet{Nakane2024b,Nakane2025} for more details of the models. 
%We treat the UV-based stellar metallicity $Z_*$ as a proxy for stellar Fe/H, as in previous studies (e.g., \citealt{Steidel2016,Cullen2019,Nakane2024b,Nakane2025}). 
Based on Equations \ref{eq:PCM}$-$\ref{eq:tau}, we set the following free parameters: a common velocity-independent covering fraction $C_f$, absorption line parameters (column density $\log N$, velocity $v_0$, and Doppler width $b_D$ for each ion), and continuum parameters (stellar metallicity $Z_*$, stellar age $\log t$, color excess $E(B-V)$, normalization factor $f_\mathrm{norm}$), and O \textsc{iii}] $\lambda\lambda1661,1666$ line fluxes $F_\mathrm{OIII]1661},F_\mathrm{OIII]1666}$ if detected. We conduct MCMC fitting using \texttt{emcee} with flat priors for all the free parameters and determine the best-fit parameters and their $1\sigma$ uncertainties, as described in \ref{subsubsec:z_sys}. In Figure \ref{fig:PCM}, we present an example of the best-fit model and posterior distributions for EGS-z7p2. The best-fit model reproduces the observed absorption profiles reasonably well. The column densities of the ion with multiple detected transitions or weak lines (e.g., Fe \textsc{ii} and Al \textsc{ii}) are well constrained, while ions with only saturated transitions (e.g., C \textsc{ii}) exhibit relatively larger uncertainties\red{, demonstrating that our forward-modeling approach appropriately captures the optical-depth dependence of the column-density constraints}. We summarize the PCM fitting results in Table \ref{tab:absorption}. We obtain moderate covering fractions of $C_f\sim0.5-0.7$. Our $C_f$ measurements of three galaxies, Gz9p3, GLASS-100003, and GLASS-10021 are consistent within uncertainties with previous measurements for SPURS-A2744-7, SPURS-A2744-17, and SPURS-A2744-24, respectively \citep{VasanGC2026}.

\subsubsection{Nebular Modeling}
\label{subsubsec:nebular}

\input{Table/table_nebular}

\begin{figure*}[ht!]
    \centering
    \includegraphics[width=0.99\linewidth]{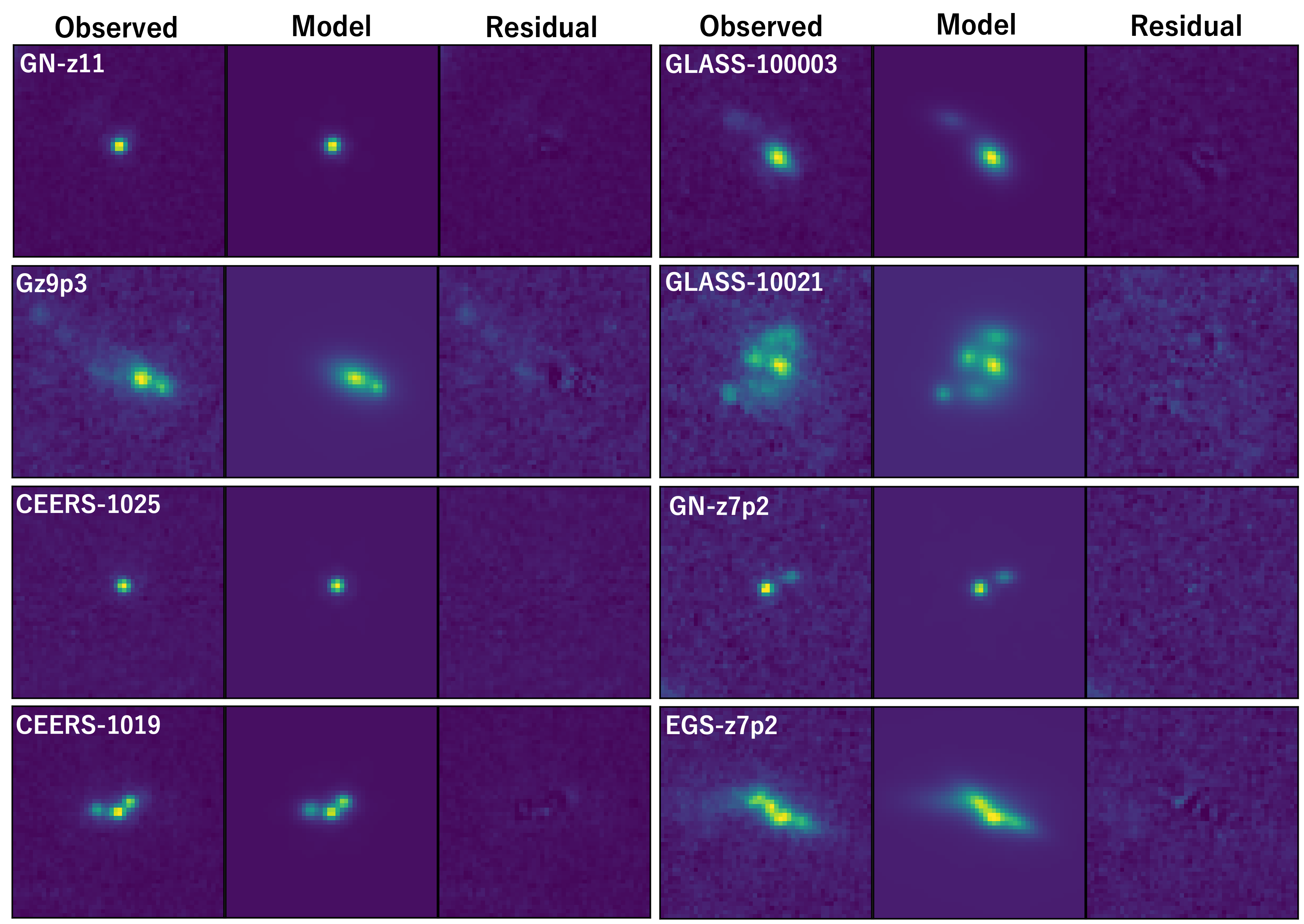}
    \caption{Morphology fitting results. In each set of three panels, the left, middle, and right panels represent the observed F200W maps, best-fit models, and residual maps, respectively}
    \label{fig:morphology}
\end{figure*}

For emission lines, we detect multiple UV and optical lines in our sample galaxies (see Table \ref{tab:line}). The optical emission lines especially trace physical properties in H \textsc{ii} regions, including gas-phase metallicity $12+\log(\mathrm{O/H})$, dust extinction $E(B-V)$, electron density $n_e$, and electron temperature $T_e$. We estimate nebular metallicities and physical conditions by forward modeling the observed emission-line spectra of [O \textsc{ii}] $\lambda\lambda3727,3729$, H$\delta$, H$\gamma$, [O \textsc{iii}] $\lambda4363$, H$\beta$, and [O \textsc{iii}] $\lambda\lambda4959,5007$. Since the H$\beta$ and [O \textsc{iii}] $\lambda\lambda4959,5007$ lines are beyond the NIRSpec wavelength coverage for GN-z11, we exclude it from the nebular modeling and adopt literature measurements (e.g., $12+\log(\mathrm{O/H})=7.91\pm0.07$; \citealt{Alvarez-Marquez2025}). The model spectra consist of a power-law continuum $\alpha\lambda^{\beta}$ and Gaussian functions for emission lines, where we fix the redshift and line width to those of [O \textsc{iii}] $\lambda4959,5007$. Since the oxygen emission lines are mostly sensitive to the gas-phase metallicity and electron temperature, the treatment of oxygen lines depends on the detections of temperature-sensitive auroral line of [O\textsc{iii}] $\lambda4363$. In both cases, we include six baseline parameters of $12+\log(\mathrm{O/H})$, $E(B-V)$, $n_e$, $F_\mathrm{H\beta}$, $\alpha$, and $\beta$. \red{We first correct $F_\mathrm{H\beta}$ for dust attenuation to obtain the intrinsic H$\beta$ flux, assuming the \citet{Calzetti2000} attenuation law. We then calculate the intrinsic line emissivities as functions of electron temperature and density using \texttt{PyNeb} \citep{Luridiana2015}, and use these emissivities together with the relevant ionic abundances or line-ratio parameters to predict the intrinsic emission-line fluxes. We finally attenuate the intrinsic fluxes to obtain the corresponding observed fluxes. The Balmer decrements are primarily sensitive to dust attenuation, while the [O \textsc{ii}] $\lambda3727$/$\lambda3729$ ratio is primarily sensitive to electron density. For galaxies without [O \textsc{iii}] $\lambda4363$ detections, we adopt an electron temperature of $T_e=\mathrm{15000} K$, which has only a minor effect because both the Balmer emissivity ratios and the [O \textsc{ii}] doublet ratio depend only weakly on $T_e$ over the relevant temperature range. We do not apply a stellar Balmer-absorption correction in our fiducial analysis, as a conservative stellar absorption correction of EW$=4$ \AA\ for H$\beta$ \citep{Curti2023} changes the inferred metallicities by $<0.1$ dex even for galaxies with relatively weak H$\beta$ emission ($\mathrm{EW}\sim20$ \AA).}

For galaxies with [O \textsc{iii}] $\lambda4363$ detected at $\mathrm{S/N}>3$ (CEERS-1019, GLASS-100003, and GLASS-10021), we add three parameters, [O \textsc{iii}] electron temperature $T_e$([O\textsc{iii}]), [O \textsc{ii}] electron temperature $T_e$([O\textsc{ii}]), and ionized oxygen number ratios $\log(\mathrm{O^+/O^{2+}})$, following the direct $T_e$ method. The [O \textsc{iii}] $\lambda\lambda4959,5007$/[O \textsc{iii}] $\lambda4363$ ratio is derived from $T_e$([O\textsc{iii}]). For the electron temperature of [O \textsc{ii}], we adopt a $T_e$([O\textsc{ii}])-$T_e$([O\textsc{iii}]) relation 
\begin{align}
    T_e([\mathrm{O}\textsc{ii}])=(0.58\pm0.19)\times T_e([\mathrm{O}\textsc{iii}])+(4520\pm2000) \mathrm{K},
\end{align}
which is calibrated with $z\sim3-5$ galaxies \citep{Chakraborty2025}. We predict the [O \textsc{iii}] $\lambda\lambda4959,5007$ and [O \textsc{ii}] $\lambda\lambda3727,3729$ fluxes by requiring consistency between the free metallicity parameter and the ionic abundances derived from the following equations
\begin{align}
    \frac{\mathrm{O}}{\mathrm{H}}&=\frac{\mathrm{O^{+}}}{\mathrm{H^+}}+\frac{\mathrm{O^{2+}}}{\mathrm{H^+}},\label{eq:direct_Te1}\\ 
    \frac{\mathrm{O^+}}
    {\mathrm{H^+}}&=\frac{F_\mathrm{[OII]3727,3729}}{F_\mathrm{H\beta}}\frac{\epsilon_\mathrm{H\beta}}{\epsilon_\mathrm{[OII]3727,3729}},\label{eq:direct_Te2}\\
    \frac{\mathrm{O^{2+}}}{\mathrm{H^+}}&=\frac{F_\mathrm{[OIII]4959,5007}}{F_\mathrm{H\beta}}\frac{\epsilon_\mathrm{H\beta}}{\epsilon_\mathrm{[OIII]4959,5007}},\label{eq:direct_Te3}
\end{align}
where we derive line emissivity with $n_e$, $T_e$([O\textsc{iii}]), and $T_e$([O\textsc{ii}]). We ignore neutral oxygen, $\mathrm{O^{3+}}$, and higher-order oxygen ions in the same way as \citet{Izotov2006}. 

For galaxies with [O \textsc{iii}] $\lambda4363$ not detected (Gz9p3, CEERS-1025, GN-z7p2, and EGS-z7p2), we introduce additional two parameters of R3 ($=F_\mathrm{[OIII]5007}/F_\mathrm{H\beta}$) and R2 ($=F_\mathrm{[OII]3727,3729}/F_\mathrm{H\beta}$) ratios, following the strong-line method. We utilize strong-line metallicity calibration based on $z\sim3-10$ galaxies \citep{Chakraborty2025}. For given metallicity values, we derive R3 and R2 values, converting them to [O \textsc{iii}] $\lambda\lambda4959,5007$ and [O \textsc{ii}] $\lambda3727,3729$ line fluxes, respectively.

We fit the model spectra to the observed spectra in the rest-frame wavelength ranges of $1250-5100$ \AA, masking other emission and absorption lines. We conduct MCMC fitting with \texttt{emcee}, adopting flat priors for $12+\log(\mathrm{O/H})$, $E(B-V)$, $n_e$, $f_\mathrm{H\beta}$, $\log\alpha$, and $\beta$. For galaxies with [O \textsc{iii}] $\lambda4363$ detections, we also use flat priors for $\log(\mathrm{O^{+}/O^{2+}})$ and $T_e$ ([O\textsc{iii}]), while we utilize a Gaussian prior for $T_e$([O\textsc{ii}]) based on the $T_e$([O\textsc{iii}])-$T_e$([O\textsc{ii}]) relation and its $1\sigma$ uncertainty of \citet{Chakraborty2025}, treating it as a latent parameter. Similarly, for galaxies without [O \textsc{iii}] $\lambda4363$ detections, we treat R3 and R2 as latent parameters and adopt Gaussian priors based on the strong-line metallicity relations and their $1\sigma$ uncertainties \citep{Chakraborty2025}. We determine the best-fit parameters and their $1\sigma$ uncertainties from the resulting posterior distributions in the same way as in Section \ref{subsubsec:z_sys}. \red{For comparison, we also derive the dust-corrected R3 and R2 ratios for the three galaxies with direct-$T_e$ measurements and estimate their strong-line metallicities using the same calibrations. The joint R2+R3 metallicities are consistent with the direct-$T_e$ measurements for all three galaxies within the uncertainties.} In Table \ref{tab:nebular}, we summarize the derived nebular properties.

\subsection{Photometric Analysis}
\label{subsec:phto}
\begin{figure}[ht!]
    \centering
    \includegraphics[width=0.99\linewidth]{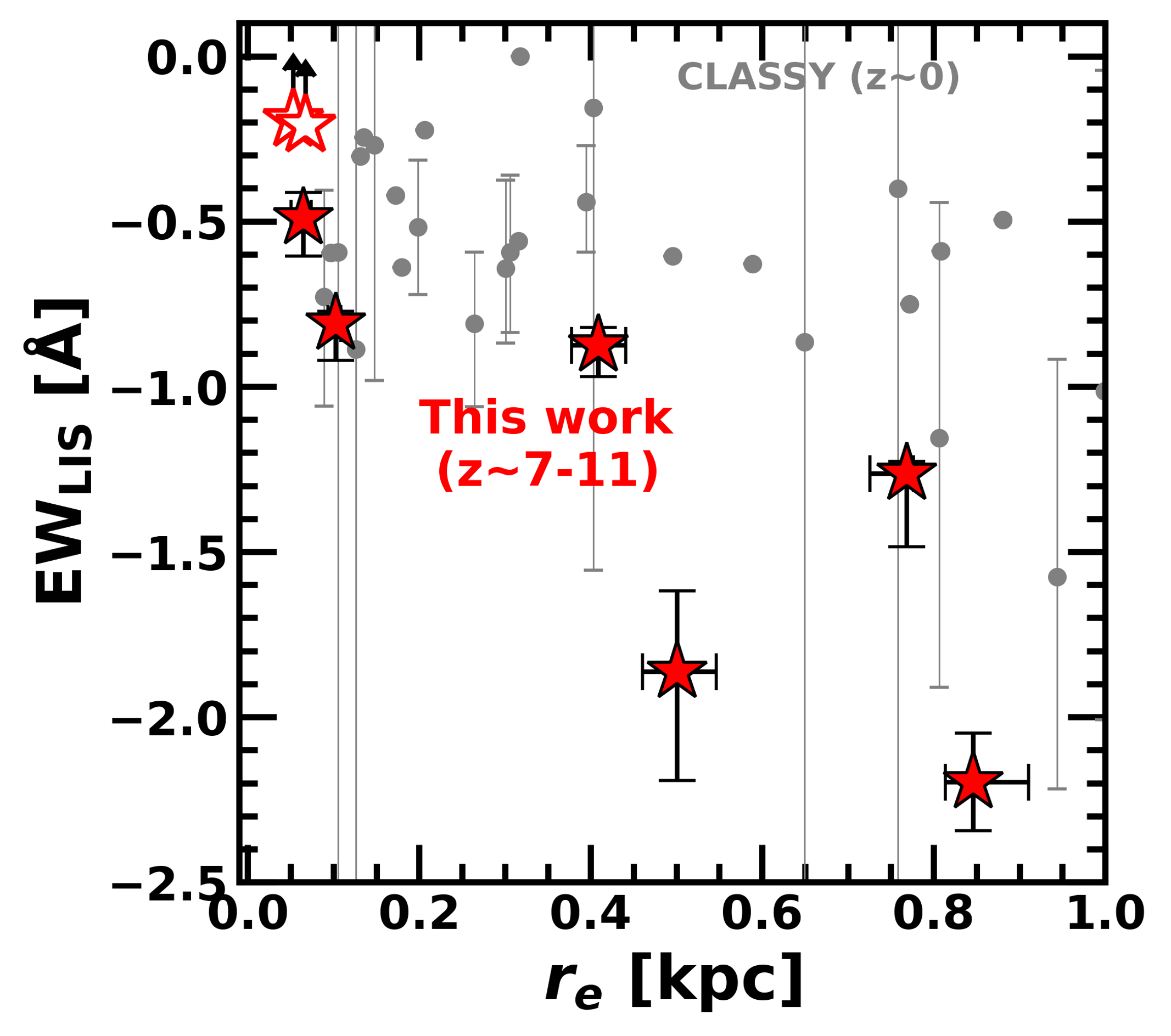}
    \caption{Average equivalent width of LIS absorption lines as a function of half-light radius. The red filled and open star symbols present measurements and $3\sigma$ lower limit for our sample galaxies at $z\sim7-11$. The gray circles show the measurements of CLASSY galaxies at $z\sim0$ \citep{Berg2022,Praker2024}.}
    \label{fig:re-EW_LIS}
\end{figure}
% \subsubsection{Morphology}
\subsubsection{\red{Size Measurement}}
\label{subsubsec:morphology}
As shown in Figure \ref{fig:sample}, our sample galaxies show diverse morphologies. \red{Here, we focus on measuring their UV sizes as a basic morphological property and investigating their relation to the LIS absorption-line strengths. We therefore} perform S\'ersic profile fitting to NIRCam/F200W map, tracing UV stellar light. We first conduct source detection and deblending using \texttt{Photutils} \red{\citep{Bradley2025}} to initially estimate the center positions and to deblend multiple components. For extended galaxies of Gz9p3, CEERS-1019, GLASS-100003, GLASS-10021, GN-z7p2, and EGS-z7p2, we detect $2$, $3$, $2$, $5$, $2$, and $4$ components, respectively, based on initial segmentation maps. We generate multiple S\'ersic profile models convolved with the PSFs, setting the following free parameters for each component: center positions $x\ \mathrm{[pix]}$, $y\ \mathrm{[pix]}$, magnitude $m_\mathrm{UV}\ \mathrm{[mag]}$, half-light radius $r_e\ \mathrm{[pix]}$, axis ratio $q$, and position angle PA [$^\circ$]. To reduce parameter degeneracies, we fix a S\'ersic index to $n=1.0$, which is a typical value for high-$z$ galaxies (e.g., \citealt{Ono2025}). \red{We confirm that varying the S\'ersic index over a reasonable range does not affect the compact/extended classification and the $r_e$-EW$_\mathrm{LIS}$ relation.} To obtain posterior distributions of the following free parameters, we conduct MCMC fitting with \texttt{emcee}, adopting flat priors for the free parameters, $-3<x<3$, $-3<y<3$ (for initial values estimated from \texttt{Photutils}), $23<m_\mathrm{UV}<30$, $0.1<r_e<10.0$, $0.1<q<1.0$, and $0<\mathrm{PA}<180$. For galaxies with multiple components (e.g., GLASS-100003), we adjust the parameter bounds to achieve stable fits. Figure \ref{fig:morphology} presents our best-fit models. From the best-fit F200W magnitudes, we calculate absolute UV magnitudes presented in Figure \ref{fig:z-M_UV} and Table \ref{tab:sample}. We derive the galaxy half-light radius from the curve of growth of the best-fit intrinsic surface brightness profile, defined as the radius enclosing half of the total intrinsic flux from all fitted components. Note that for CEERS-1019, GLASS-100003, and GN-z7p2, we adopt the effective radius of the brightest UV clump as a representative size because it dominates the UV luminosity and is separated from the other clumps in the F200W map. We present the $r_e$ measurements in Table \ref{tab:sample}. Four galaxies of GN-z11, CEERS-1025, CEERS-1019, and GN-z7p2 shows compact sizes ($r_e\lesssim100$ pc) while the remaining four galaxies of Gz9p3, GLASS-100003, GLASS-10021, and EGS-z7p2 exhibit larger sizes ($r_e\gtrsim400$ pc).

\subsubsection{SED Fitting}
\label{subsubsec:sed}

To examine the stellar populations of our sample galaxies, we conduct SED fitting to the photometry and line fluxes, using \texttt{Bagpipes} \citep{Carnall2018}. In the SED fitting, we include emission line \red{fluxes} of C \textsc{iii}] $\lambda\lambda1907,1909$, [O\textsc{ii}] $\lambda\lambda3727,3729$, H$\gamma$, H$\beta$, and [O \textsc{iii}] $\lambda\lambda4959,5007$, wherever available. To ensure consistency with the photometry, we scale the spectroscopic line fluxes (Section \ref{subsubsec:line_flux}) by a single factor $f_\mathrm{scale}$ for each galaxy. We determine $f_\mathrm{scale}$ by comparing the HST/JWST photometry and filter-convolved NIRSpec spectra at rest-frame wavelengths longer than $1220$ \AA\ to avoid the Lyman break.  We adopt stellar population synthesis models of \citet{Bruzual2003} and include nebular emission calculated with \texttt{Cloudy}. We fix the redshifts to the systemic redshifts (Section \ref{subsubsec:z_sys}) and assume a \citet{Calzetti2000} attenuation law. We adopt a non-parametric SFH consisting of seven time bins: $0-5$, $5-10$, $10-30$, $30-50$, $50-100$, $100-300$, and $300-t_\mathrm{age}$ Myr. The model has $10$ free parameters of the formed stellar mass, formed stellar metallicity, dust attenuation, ionization parameter, and six adjacent-bin SFR ratios. We adopt flat priors of $5<\log(M_\mathrm{form}/M_\odot)<13$, $0.01<Z/Z_\odot<1$, $0<A_\mathrm{V}<3$, $-3<\log U<-0.5$, and $-3<\log(\mathrm{SFR}_{i+1}/\mathrm{SFR}_{i})<3$ ($i=1-6)$. Table \ref{tab:SED} summarizes the stellar populations derived from the SED fitting.
\input{Table/table_SED}

\section{Results}
\label{sec:result}
\subsection{Absorption Line Strengths}
\label{subsec:abs_line_property}
We compare our equivalent width measurements with previous studies for galaxies in common samples. For GLASS-100003 and GLASS-10021, our measurements are generally consistent with those reported for Galaxy B and Galaxy C in \citet{Zhu2026}, respectively. For Gz9p3, our measurements are slightly lower than those reported for SPURS-A2744-7 in \citealt{Chen2026a} and Galaxy A in \citealt{Zhu2026}, but remain consistent within uncertainties, likely reflecting differences in continuum level determinations. These \red{comparisons} confirm the robustness of our EW measurements. We derive mean EWs of LIS absorption lines $\mathrm{EW_{LIS}}$ using relatively strong, unblended transitions of Si \textsc{ii} $\lambda\lambda1260,1526$ and C \textsc{ii} $\lambda1334$. As presented in Table \ref{tab:line}, our measurements span a wide range of $>-0.19$ \AA\ to $-2.20$ \AA. For comparison, a recent stacking analysis of $z=6.0-9.4$ galaxies from JWST/NIRSpec medium-resolution spectra yields $\mathrm{EW_{LIS}}=-1.18_{-0.28}^{+0.26}$ \AA \ \citep{Glazer2025}. This value based on the stacked spectra lies within the ranges of our individual measurements, but does not capture the full diversity observed among individual galaxies. We note that although the $\mathrm{EW_{LIS}}$ measurement of the stacked spectra includes O \textsc{i} $\lambda1302$ and Si \textsc{ii} $\lambda1304$ in addition to the absorption lines used in our analysis, this difference does not affect our conclusion. In Figure \ref{fig:re-EW_LIS}, we present $\mathrm{EW_{LIS}}$ as a function of effective radius of galaxies. We find a possible trend between galaxy size and LIS EWs among our sample galaxies. Compact galaxies such as GN-z11 ($r_e=68$ pc) and CEERS-1025 ($r_e=53$ pc) show no significant LIS absorption, while more extended galaxies, including GLASS-10021 ($r_e=769$ pc) and EGS-z7p2 ($r_e=846$ pc), exhibit strong absorption features. We compare our sample galaxies with local UV-bright galaxies from the Cosmic Origins Spectrograph Legacy Spectroscopic Survey (CLASSY; \citealt{Berg2022}). We take the $r_e$ measurements from \citet{Berg2022} and calculate $\mathrm{EW_{LIS}}$ from measurements of the same LIS lines obtained in \citet{Praker2024}. The CLASSY galaxies shows a similar but weak trend between $r_e$ and $\mathrm{EW_{LIS}}$. Although the scatter is large, the local CLASSY galaxies follow a qualitatively similar trend, suggesting that the connection between galaxy size and LIS absorption strength may persist over a wide range of redshift. We further discuss the origins of correlation between the galaxy sizes and strength of LIS absorption lines in Section \ref{subsec:weak_LIS}.

\subsection{Abundance Ratios}
\label{subsec:abundance_ratios}
\begin{figure*}
    \centering
    \includegraphics[width=0.99\linewidth]{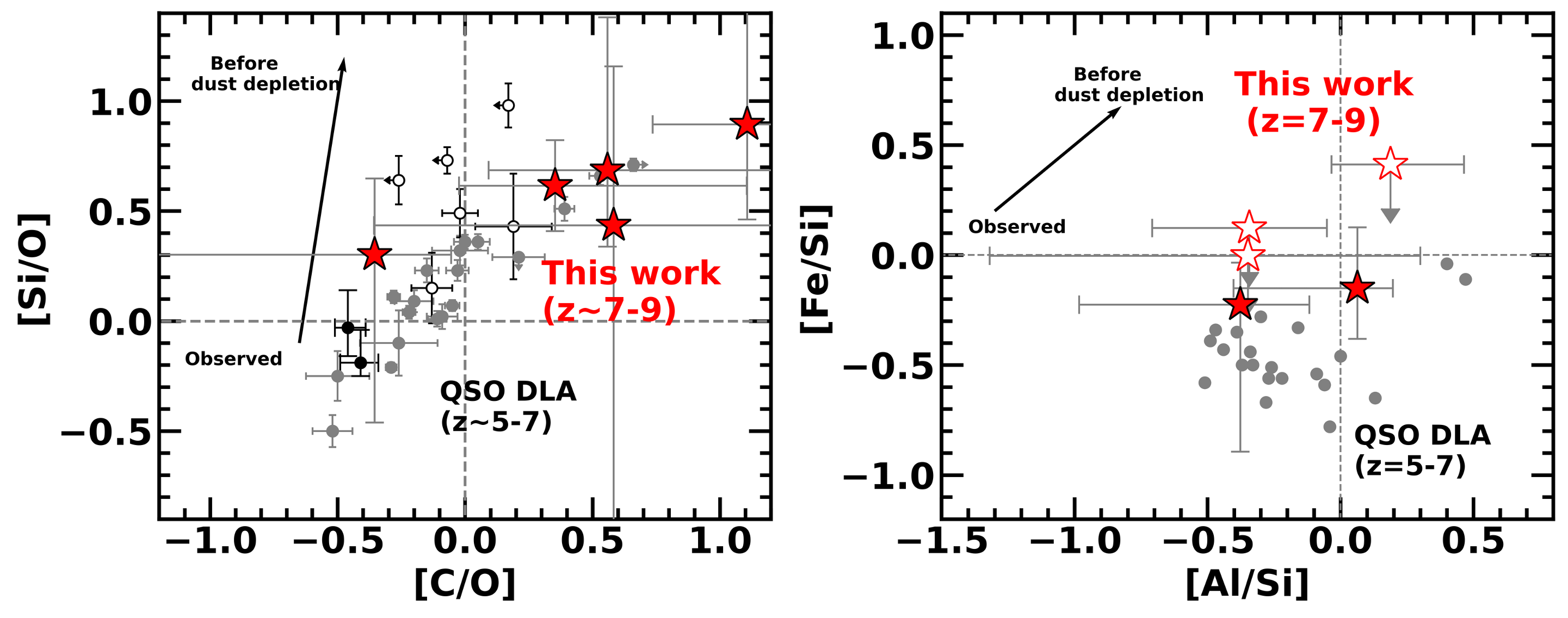}
    \caption{Abundance ratios of [Si/O] and [C/O] (left), and [Fe/Si] and [Al/Si] (right). The red filled and open star symbols present measurements and $3\sigma$ upper limit for our sample galaxies. The gray and white circles denote the measurements of QSO DLA systems \citep{Christensen2023,Sodini2024} while the black circles show emission-line based measurements of galaxies at $z>6$ \citep{Isobe2026}. The black arrows indicate the effects of dust depletion based on depletion factor defined by \citet{Ferland2013}.}
    \label{fig:abundance}
\end{figure*}

\input{Table/table_abundance}
We derive abundance ratios of LIS gas from the ionic column densities obtained in Section \ref{subsubsec:abs_model}. We do not apply ionization corrections, assuming that the LIS absorption gas is predominantly neutral, as commonly assumed for LIS absorption systems (e.g., \citealt{Jones2018,Sodini2024}). 
%v1.5
We summarize the abundance ratio measurements in Table \ref{tab:abundance}.
In the left panel of Figure \ref{fig:abundance}, we compare the [C/O] and [Si/O] abundance ratios of our sample galaxies with those of QSO DLA systems \red{at $z\sim5-7$} \citep{Christensen2023,Sodini2024} and individual galaxies at $z>6$ \citep{Isobe2026}. Our sample galaxies exhibit high (possibly supersolar) [C/O] and [Si/O] ratios although some DLA measurements reach similarly high values. The high [C/O] measurements in cool gas for Gz9p3, CEERS-1019, and GLASS-100003 are comparable to those reported for Gz9p3, EGSY8p7, and ZD2 in \citealt{Pollock2026}, respectively, while [Si/O] values are higher than their measurements. The differences in dust depletion, depending on elements, may affect the abundance ratios. If dust depletion is significant, the intrinsic total (gas + dust) abundance ratios would be \red{shifted relative to the observed gas-phase values, as indicated by the arrow in the figure. The direction of the arrow represents the relative depletion between the elements, while its length indicates the approximate magnitude of the correction \citep{Ferland2013}.} The right panel shows [Fe/Si] and [Al/Si] abundance ratios. Among the five galaxies with abundance constraints, only GLASS-10021 and EGS-z7p2 show Fe \textsc{ii} detections, enabling direct constraints on iron abundance ratios. Both galaxies have near-solar abundance ratios of $\mathrm{[Fe/Si]}\gtrsim-0.2$, placing them at the upper end of the DLA distribution. These values are comparable to those observed in Milky Way (MW) stars with metallcities above $\gtrsim30$\% solar (e.g., \citealt{Bensby2013}), suggesting that some degree of Fe enrichment has already occurred in these galaxies. We discuss the implications of those abundance patterns for the origin of Fe enrichment in Section \ref{subsec:Fe_enrichment}.

\section{Discussion}
\label{sec:discussion}
\subsection{Weak LIS Absorption Lines}
\label{subsec:weak_LIS}

\begin{figure}[ht!]
    \centering
    \includegraphics[width=0.99\linewidth]{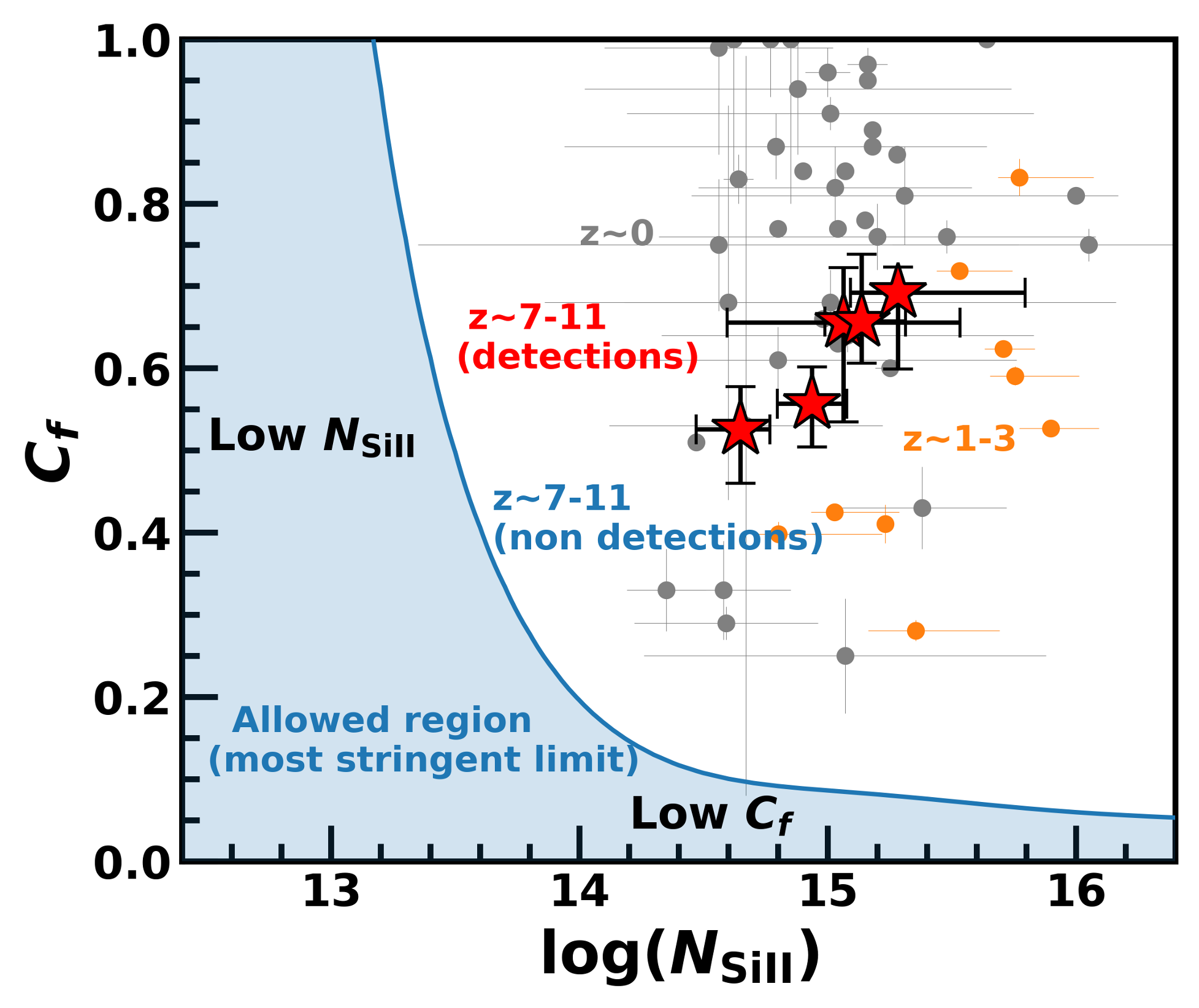}
    \caption{Covering fraction as a function of Si \textsc{ii} column density. The red star symbols denote the measurements of our sample galaxies at $z\sim7-11$ with Si \textsc{ii} lines detected. The blue lines and shaded region show the most stringent limit of $\mathrm{EW_{SiII}}>-0.2$ \AA\ for our sample galaxies at $z\sim7-11$ with Si \textsc{ii} lines non-detected and allowed region by the limit, respectively. The orange and gray circles are the measurement of $z\sim1-3$ star-forming galaxies \citep{Jones2018} and $z\sim0$ CLASSY galaxies \citep{Praker2024}, respectively.}
    \label{fig:N-Cf}
\end{figure}

We find a possible correlation between LIS EWs and $r_e$ for local and high-$z$ galaxies in Figure \ref{fig:re-EW_LIS}. Particularly, high-$z$ compact galaxies ($r_e\lesssim100$ pc) show weak or absent LIS absorption features. Recent works by \citet{Glazer2025,Chen2026a} also report weak LIS absorptions of high-$z$ galaxies, which may be due to low LIS covering fraction. However, EWs of absorption lines depend not only on the covering fraction, but also on the column densities. To explore these effects, we compare galaxies with and without absorption features on the $N$-$C_f$ plane in Figure \ref{fig:N-Cf}. Our sample galaxies with Si \textsc{ii} detections show comparable Si \textsc{ii} column densities and covering fractions to those of galaxies at $z\sim0$ \citep{Praker2024} and $z\sim2-3$ \citep{Jones2018} with no significant redshift evolution. On the other hand, the blue-shaded region presents the allowed parameter space by most stringent Si \textsc{ii} $\lambda1260$ EW limit ($\mathrm{EW}>-0.2$ \AA), which suggests a low Si \textsc{ii} column density of $\log\mathrm{N_{SiII}}\lesssim13.5$ or a low covering fraction of $C_f<0.1$. \red{Similar trends are also found for O \textsc{i}, C \textsc{ii}, and Al \textsc{ii} (see Figure \ref{fig:N-Cf_appendix}), indicating that this result is not specific to Si \textsc{ii}.} Although it is difficult to distinguish between these possibilities with the current data, we find that both inferred low $N$ and $C_f$ values are significantly lower than those typically observed in local and high-$z$ galaxies. This implies that LIS gas of high-$z$ compact galaxies may be in a distinct condition compared to more extended galaxies.

\begin{figure*}[ht!]
    \centering
    \includegraphics[width=0.8\linewidth]{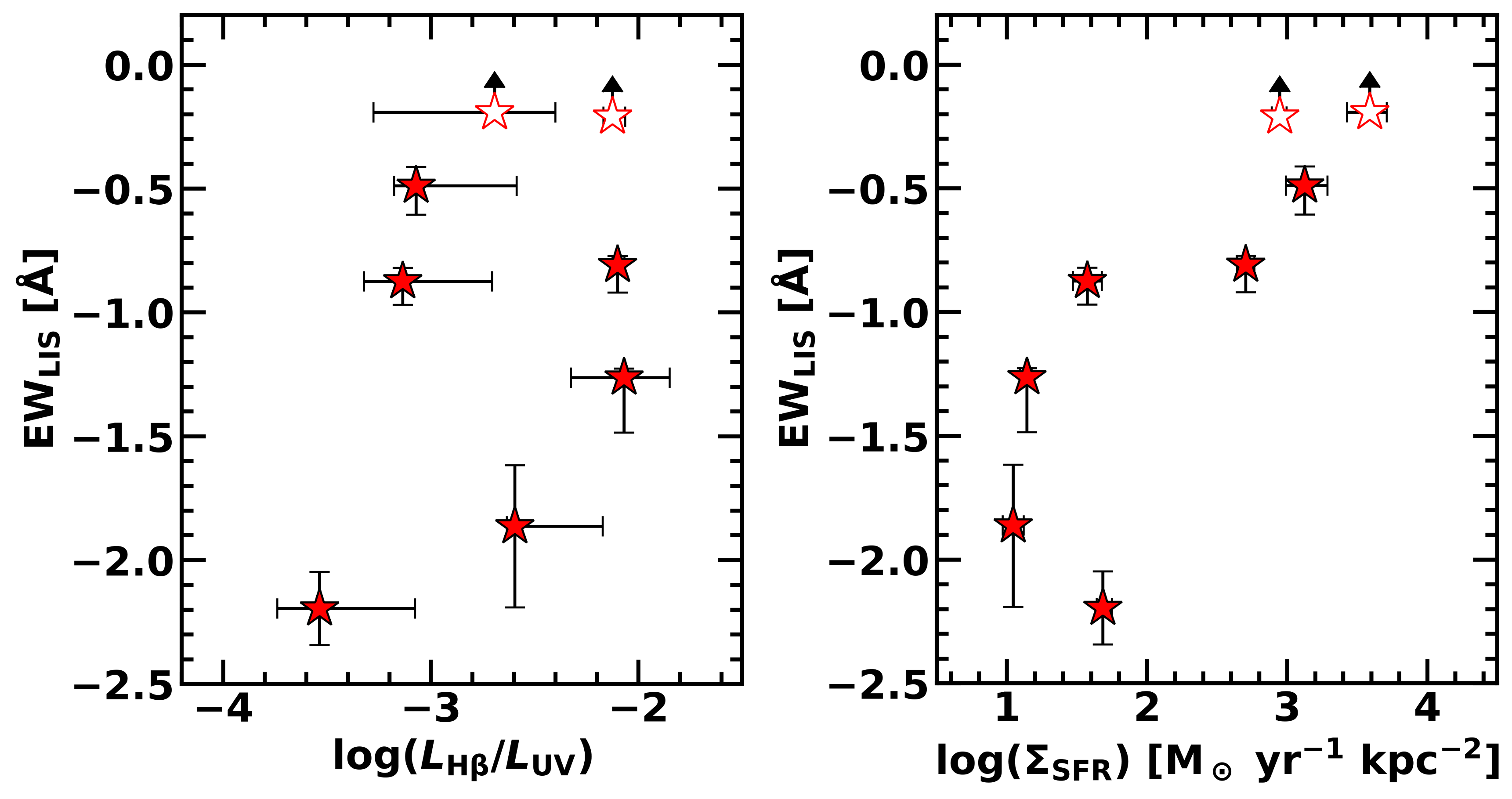}
    \caption{Same as Figure \ref{fig:re-EW_LIS}, but as a function of H$\beta$-to-UV luminosity ratio (left) and star formation rate surface density (right).}
    \label{fig:burstiness}
\end{figure*}

High-$z$ compact galaxies indeed show unique features such as bursty star formation, high-ionization emission lines, and high electron density (e.g., \citealt{Bunker2023,Castellano2024,Maiolino2024,Schaerer2024,Topping2024,Harikane2025,Li2025}). To quantify the effects of burstiness and compactness on the LIS absorptions, we derive H$\beta$-to-UV luminosity ratios $L_\mathrm{H\beta}/L_\mathrm{UV}$, \red{which trace recent star-formation burstiness in the same manner as the commonly used H$\alpha$-to-UV ratio (e.g., \citealt{Atek2022,Endsley2025}),} and SFR surface densities $\Sigma_\mathrm{SFR}$. We obtain $L_\mathrm{H\beta}/L_\mathrm{UV}$ from dust-corrected H$\beta$ and UV luminosities. \red{The H$\beta$ luminosity is corrected using the nebular attenuation inferred from the Balmer decrement (Section \ref{subsubsec:nebular}), while the UV luminosity is corrected using he stellar attenuation inferred from the SED fitting (Section \ref{subsubsec:sed}).} We estimate $\Sigma_\mathrm{SFR}$ by combining the $r_e$ measurements (Section \ref{subsubsec:morphology}) and H$\beta$-based SFRs, assuming a $L_\mathrm{H\alpha}$-$L_\mathrm{UV}$ relation \citep{Kennicutt1998} and Case B recombination. As shown in Figure \ref{fig:burstiness}, $\mathrm{EW_{LIS}}$ tends to decrease with increasing $L_\mathrm{H\beta}/L_\mathrm{UV}$ and $\Sigma_\mathrm{SFR}$. 
These results suggest two physical interpretations for the weak LIS absorption in the compact galaxies. On one hand, the high $L_{H\beta}/L_\mathrm{UV}$ values indicate a hard radiation field, which could convert low-ionization gas into high-ionization state, leading to low LIS column density. This is consistent with observations that high-$z$ compact galaxies exhibit strong high-ionization emission lines such as C \textsc{iv} and N \textsc{iv} (e.g., \citealt{Schaerer2024,Topping2024,Harikane2025}). Compact galaxies in our sample (GN-z11, CEERS-1025, CEERS-1019, and GN-z7p2) indeed exhibit the UV high-ionization lines of He \textsc{ii}, C \textsc{iv}, and N \textsc{iv} as well as \red{stellar wind signatures} (P-Cygni profiles of N \textsc{v}, C \textsc{iv}, and Si \textsc{iv}), which are driven by young stellar populations like Wolf-Rayet stars and very massive stars \citep{Chen2026b,Marques-Chaves2026,Nakane2026}. In addition, some galaxies show possible AGN signatures, including broad line H$\beta$ emission in CEERS-1019 \citep{Larson2023}. On the other hand, high star formation rate densities are expected to drive strong outflows (e.g., \citealt{Ferrara2023,Ferrara2024}), potentially expelling cool gas and reducing its covering fraction. A low LIS covering fraction is often associated with Lyman continuum (LyC) leakers, since the LIS covering fraction has been shown to be positively correlate with the H \textsc{i} covering fraction, which may regulate the escape of ionizing photon (e.g., \citealt{Chisholm2018}). This interpretation is qualitatively consistent with the strong Ly$\alpha$ emission in our sample of compact galaxies (\citealt{Bunker2023,Larson2023,Tang2025,Marques-Chaves2026}, see also Figure \ref{fig:sample}). By contrast, extended galaxies tend to exhibit weak or absent Ly$\alpha$ emission and local DLA absorption \citep{Chen2026a}. \red{The outflow scenario is also consistent with the low dust extinction of $E(B-V)\sim0$ and blue UV slope of $\beta\sim-2.4$ observed in the compact galaxies  (\citealt{Bunker2023}; see also Table \ref{tab:nebular}, except for CEERS-1019, which may have an AGN contribution.} To summarize, both enhanced ionization and feedback-driven outflows likely contribute to the weak LIS absorption in UV-bright compact galaxies at $z>7$, possibly driven by intense star formation or nuclear activity.

\subsection{Origin of Iron Enrichment}
\label{subsec:Fe_enrichment}

Abundance ratios of $\alpha$ element to iron [$\alpha$/Fe] (e.g., [O/Fe], [Si/Fe]) has been used as \red{a} ``cosmic clock" since the iron enrichment traces star formation history, as suggested by observations of the MW stars (e.g., \citealt{Bensby2013,Zhao2016}). Instantaneous core-collapse supernovae (CCSNe) by massive stars eject a lot of $\alpha$ elements relative to iron, resulting in high [$\alpha$/Fe] ratios at low \red{metallicities}. Delayed Type-Ia supernovae (SNe Ia) caused by gas accretion/binary merger of white dwarfs eject a lot of iron relative to $\alpha$ elements, which reduces [$\alpha$/Fe] ratios. The delay time of SNe Ia is typically $0.1-1.0$ Gyr, requiring long star formation history. At $z\sim2-6$, Fe abundance measurement based on Fe photospheric lines in the UV stellar continuum from the stacked galaxy spectra indicates $\alpha$ enhancement (\red{$\mathrm{[O/Fe]}>0$}), which is consistent with enrichment by CCSNe (e.g., \citealt{Steidel2016,Cullen2019,Harikane2020,Cullen2021,Kashino2022}). However, recent works report substantial Fe enrichment (\red{$\mathrm{[O/Fe]}<0$}) for galaxies at $z\gtrsim6$ (corresponding to less than 1 Gyr in cosmic age) based on Fe emission lines from the individual galaxy \citep{Ji2024a,Ji2024b,Tacchella2025} and stacked galaxy spectra \citep{Isobe2026}, Fe photospheric lines in the UV stellar continuum from the individual galaxy spectra \citep{Nakane2024b,Nakane2025}. The observations of QSOs at $z\sim3-7$ has also identified signatures of Fe enhancement based on high Fe \textsc{ii}/Mg \textsc{ii} ratios (e.g., \citealt{DeRosa2011,Sameshima2017,Onoue2020,Sameshima2020}). This emerging picture contrasts with the $\alpha$-enhanced abundance patterns observed in typical star-forming galaxies at $z~\sim2-6$. Without considering the short delay time, it may be difficult to explain the Fe enhancement by SNe Ia as early as $z\gtrsim6$. This has motivated alternative explanations for the apparent Fe enrichment, including enrichment by hypernovae (HNe) and theoretical pair-instability supernovae (PISNe), whose large explosion energies lead to destruction of the inner core of massive stars, ejecting a lot of iron (e.g., \citealt{Nomoto2013,Takahashi2018}).

Previous work  shows that Fe-rich abundance ratios of $z>10$ galaxies could be reproduced either by SNe Ia with short delay times ($\sim30-50$ Myr) or by PISNe with \red{a lower contribution from CCSNe, based on comparisons} with chemical evolution models \citep{Nakane2025}. Abundance ratios involving iron alone are often insufficient to distinguish between these scenarios. One powerful diagnostic is the odd-even effect (e.g., \citealt{Nomoto2013}), which reflects differences in neutron excess during nucleosynthesis. Since PISNe originate from metal-poor very massive stars, they are expected to have a low neutron excess compared to CCSNe. As a result, PISNe produce significant deficiencies of \red{odd-$Z$ elements (i.e., elements with odd atomic numbers) relative to even-$Z$ elements (i.e., elements with even atomic numbers)}, resulting in low abundance ratios such as [Al/Si]. Therefore, the combination of [Fe/Si] and [Al/Si] provides a useful way to distinguish between these enrichment pathways. For our sample, we have obtained constraints of [Fe/Si] and [Al/Si] in five galaxies (Figure \ref{fig:abundance}) from LIS absorption lines. We discuss the Fe enrichment pathways by comparing these abundance ratios with chemical evolution models described below.

\begin{figure*}[ht!]
    \centering
    \includegraphics[width=0.99\linewidth]{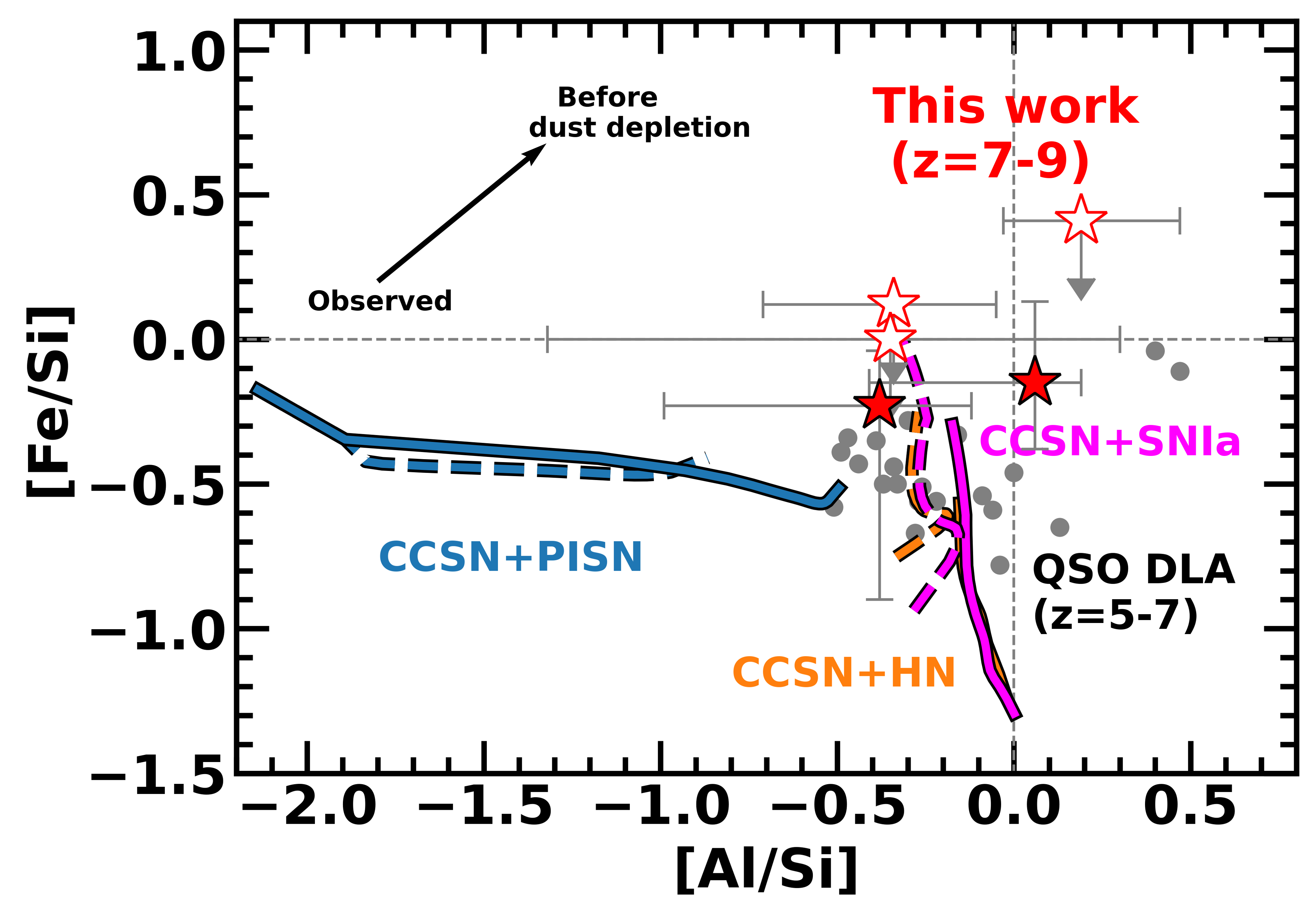}
    \caption{Comparison of [Fe/Si] and [Al/Si] abundance ratios with chemical evolution models. The red star symbols, gray circles, and black arrows are the same as in Figure \ref{fig:abundance}. \red{The magenta, orange, and blue lines present our SN Ia, HN, and PISN models, respectively.} The solid and dashed lines indicate models with and without failed supernovae for stars with masses of $30-100\ M_\odot$.}
    \label{fig:AlSi-FeSi}
\end{figure*}

We construct chemical evolution models in the same way as \citet{Nakane2025}. The models have \red{three} sets of yields, CCSN + SN Ia yields (SN Ia models), CCSN + PISN yields (PISN models), \red{and CCSN + HN yields (HN models)}. We assume a \citet{Salpeter1955} IMF and a constant SFH with stellar ages of $0-200$ Myr. We use yield models of CCSNe \red{and HNe} \citep{Nomoto2013}, PISNe \citep{Takahashi2018}, and SNe Ia \citep{Iwamoto1999} and \red{stellar} lifetimes derived by \citet{Padovani1993}. We calculate the IMF-averaged yields, and sum them according to the \red{stellar} lifetimes, where we assume CCSNe, \red{HNe,} PISNe, and SNe Ia are caused by stars with masses of $9-100\ M_\odot$, \red{$20-40\ M_\odot$,} $140-300\ M_\odot$, and $1-8\ M_\odot$. \red{For the HN models, we adopt an HN fraction of $f_\mathrm{HN}=0.5$ for stars with masses of $20-40\ M_\odot$. This corresponds to the high HN fraction adopted for very metal-poor stellar populations in the metallicity-dependent prescription of \citet{Kobayashi2020}, and therefore provides a conservative test of the maximum plausible HN contribution.} For SNe Ia, we consider the delay time distribution (DTD) expressed by a power-law function (e.g., \citealt{Maoz2014}) with a fiducial delay time of $100$ Myr. In the SN Ia models, we also include the enrichment from asymptotic giant branch (AGB) stars with initial masses of $1-8$ $M_\odot$, using the yields of \citet{Karakas2010}. We finally obtain abundance ratios by converting the yield mass ratios to number ratios. For comparison, we include the models with failed SNe for stars with masses of $30-100\ M_\odot$, which are suggested as sources of reducing contributions from CCSNe to explain high N/O and Fe/O ratios in recent studies (e.g., \citealt{Watanabe2024,Nakane2025}).

In Figure \ref{fig:AlSi-FeSi}, we compare our [Fe/Si] and [Al/Si] measurements with our chemical evolution models. The observed abundance ratios are inconsistent with the PISN models, which predict significantly lower [Al/Si] ratios owing to the strong odd-even effect. This discrepancy remains even after including the CCSN contribution and suggests that PISNe are unlikely to be the dominant source of the observed Fe enrichment. 
\red{While near-solar [Al/Si] ratios can be produced by CCSNe alone, the near-solar [Fe/Si] ratios require an additional source of iron, naturally explained by SNe Ia. The HN models without failed SNe remain relatively Fe poor, while the HN models including failed SNe can approach the observed abundance ratios. However, this requires the high HN fraction of $f_\mathrm{HN}=0.5$, comparable to the high values adopted for very metal-poor stellar populations in \citet{Kobayashi2020}, and also depends on the assumed failed-SN prescription.}
%In contrast, the SN Ia models reproduce both the observed [Fe/Si] and [Al/Si] ratios. While near-solar [Al/Si] ratios can be produced by CCSNe alone, the near-solar [Fe/Si] ratios require an additional source of iron, naturally explained by SNe Ia. 

\begin{figure*}[ht!]
    \centering
    \includegraphics[width=0.99\linewidth]{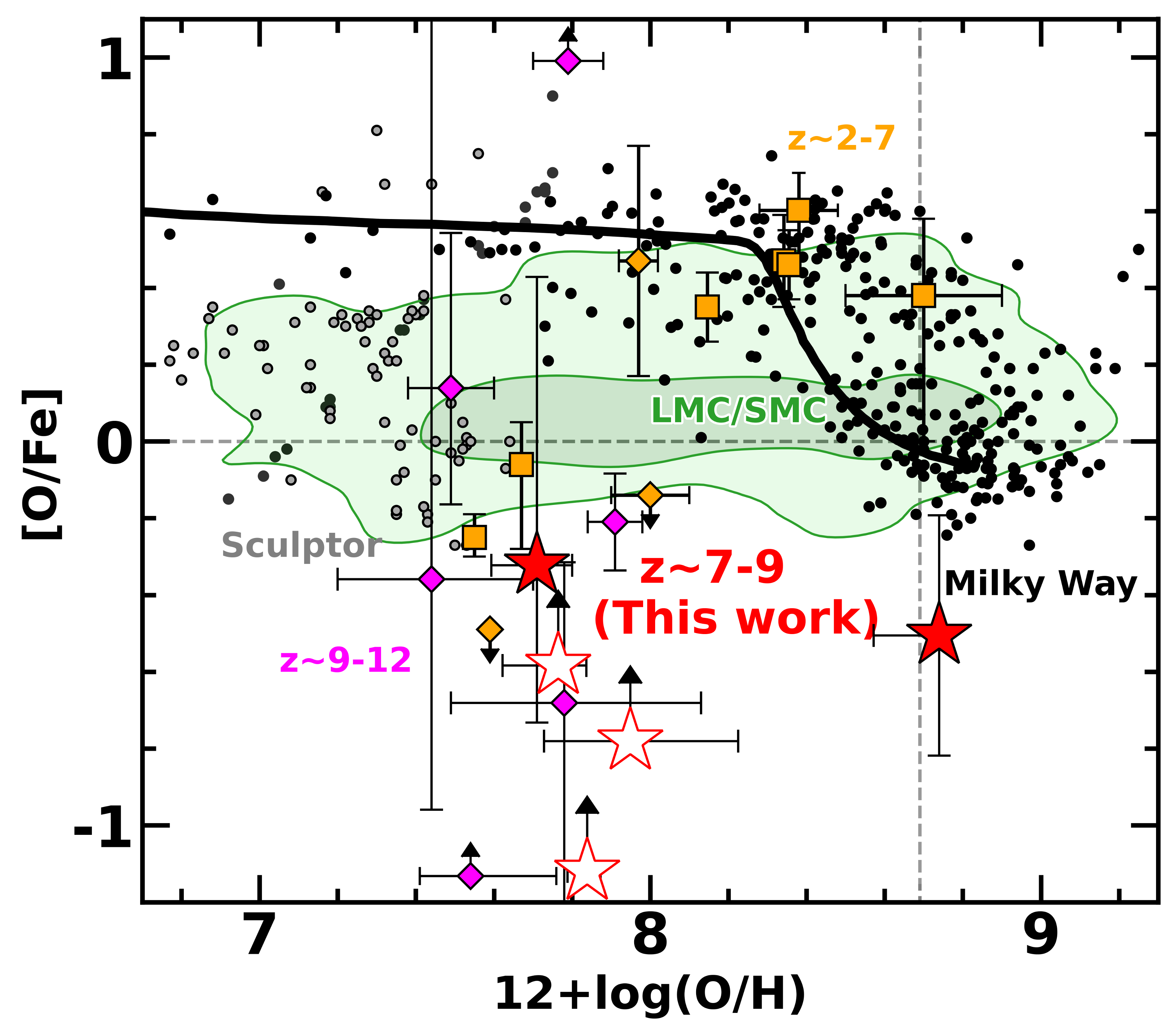}
    \caption{[O/Fe] as a function of $12+\log(\mathrm{O/H})$. The red star symbols show measurements of our sample galaxies at $z\sim7-9$ based on the UV LIS absorption lines. The magenta and orange diamonds represent the measurements of individual galaxies at $z\sim9-12$ based on UV stellar absorptions \citep{Nakane2025} and at $z\sim2-7$ based on emission lines \citep{Ji2024a,Tacchella2025,Welch2025}\red{, respectively}. The orange square denote measurements for stacked galaxy spectra at $z\sim2-7$ based on UV stellar absorptions \citep{Steidel2016,Cullen2019,Harikane2020,Cullen2021,Kashino2022} and based on emission lines \citep{Isobe2026}. The black circles, gray circles, and green contours denote the measurements of local stars in MW \citep{Melendez2003,Carretta2005,Lecureur2007,Pasquini2008,Valenti2011,Bensby2013,Zhao2016,Amarsi2019}, Sculptor galaxy \citep{Tang2023}, and LMC/SMC \citep{Hasselquist2021}, respectively. The black solid line shows the chemical evolution model, presented in Figure 15 of \citet{Kobayashi2020}. \red{See Section \ref{subsec:Fe_enrichment} for a brief description of the model.}}
    \label{fig:OH-OFe}
\end{figure*}

In Figure \ref{fig:OH-OFe}, we compare our measurements of [O/Fe] and $12+\log(\mathrm{O/H})$ with those of the local stars (MW; \citealt{Melendez2003,Carretta2005,Lecureur2007,Pasquini2008,Valenti2011,Bensby2013,Zhao2016,Amarsi2019}, LMC/SMC; \citealt{Hasselquist2021}, Sculptor galaxy; \citealt{Tang2023}) and high-$z$ galaxies \citep{Steidel2016,Cullen2019,Harikane2020,Cullen2021,Kashino2022,Ji2024a,Tacchella2025,Nakane2025,Welch2025,Isobe2026}. As described above, [O/Fe] ratios measured from stacked galaxy spectra at $z\sim2-6$ based on the stellar continuum are generally supersolar and broadly consistent with MW stars and \red{the chemical evolution model presented in Figure 15 of \citet{Kobayashi2020}, which includes metallicity-dependent yields from CCSNe and HNe, as well as enrichment from Chandrasekhar-mass SNe Ia.}  On the other hand, several galaxies at $z\gtrsim6$ show subsolar [O/Fe] ratios at comparable metallicities \citep{Ji2024a,Tacchella2025,Nakane2024b,Nakane2025}. If \red{this} Fe enrichment is mainly caused by SNe Ia, the ``knee" of the [O/Fe] evolution would occur at substantially lower metallicities than in MW. Interestingly, a similarly early decline of [O/Fe] has been observed in stars of the Sculptor dwarf galaxies \citep{Tang2023}, suggesting that rapid Fe enrichment may not be unique to the early Universe. One possible explanation is a significant contribution \red{from} sub-Chandrasekhar mass SNe Ia, which are predicted to occur earlier than Chandrasekhar mass SNe Ia and can contribute to Fe enrichment on timescales of $\si40$ Myr (e.g., \citealt{Kobayashi1998,Kobayashi2020,Bhattacharya2025}). The two Fe \textsc{ii}-detected galaxies in our sample, GLASS-10021 and EGS-z7p2, show both near-solar [Fe/Si] ratios and young mass-weighted stellar ages of only $\sim15-20$ Myr (see Table \ref{tab:SED}). If these ages reflect the dominant star-formation episode, the inferred Fe enrichment timescale is even shorter than the prompt SN Ia channel, potentially involving sub-Chandrasekhar-mass progenitors. This suggests that the early Fe enrichment would be naturally explained by prompt SNe Ia and dual starbursts (e.g., \citealt{Kobayashi&Ferrara2024}). Our results therefore favor early SN Ia enrichment as the origin of the Fe enrichment in galaxies at $z\sim7$. \red{However, this conclusion does not necessarily apply to other Fe-rich galaxies at $z\gtrsim6$, for which the odd-even abundance pattern is not yet well constrained and PISN enrichment may still provide a viable explanation.} 

%\red{Finally, the UV morphologies qualitatively reinforce the $r_e$-$\mathrm{EW_{LIS}}$ relation discussed above. Galaxies dominated by a single compact UV component show weak or undetected LIS absorption, whereas galaxies with multiple UV components generally exhibit stronger and/or multiple LIS absorption lines. In particular, GLASS-10021 and EGS-z7p2, among the most extended and clump-rich galaxies in our sample, show the strongest LIS absorption and significant Fe \textsc{ii} detections. However, we find no clear trend between morphology and gas-phase metallicity, with EGS-z7p2 being the only notably metal-rich galaxy. This suggests that the morphology-LIS relation is unlikely to be driven simply by overall metallicity and may instead reflect differences in the column density and/or covering fraction of low-ionization gas. Although the sample size is small, this contrast may suggest that strong LIS absorption and Fe \textsc{ii} enrichment, as indicated by near-solar Fe/Si ratios, preferentially emerge in systems that have undergone substantial structural assembly.}

\red{Finally, we note that two Fe \textsc{ii} detected galaxies, GLASS-10021 and EGS-z7p2, are among the most extended and clumpy systems in our sample, consisting of multiple UV components, while galaxies dominated by a single compact UV component show weak or undetected LIS absorption and no significant Fe \textsc{ii} detections. We find no clear trend between UV morphology and gas-phase metallicity. Although the sample size is small, this contrast may suggest that strong LIS absorption and detectable Fe \textsc{ii} enrichment preferentially emerges in galaxies with more complex, multi-component UV morphologies, possibly associated with substantial structural assembly.}

\section{Summary} 
\label{sec:summary}
In this paper, we investigate UV LIS absorption lines and chemical enrichment of cool gas in eight UV-bright galaxies at $z=7.2-10.6$, using JWST/NIRSpec deep medium-resolution grating spectra. We detect multiple LIS lines in five galaxies\red{, including Si \textsc{ii} $\lambda\lambda1260,1304,1526$, O \textsc{i} $\lambda1302$, C \textsc{ii} $\lambda1334$, Fe \textsc{ii} $\lambda\lambda1608,2600$, Al \textsc{ii} $\lambda1670$, and Mg \textsc{ii} $\lambda\lambda2796,2803$,} while the remaining three galaxies show only one or no significant LIS line detections. We summarize our major findings below:
\begin{itemize}
    \item[1.] 
    We find a decreasing trend of mean LIS line equivalent width with increasing galaxy half-light radius for our sample galaxies at \red{$z\sim7-11$} and for literature CLASSY galaxies at $z\sim0$. Particularly, compact galaxies ($r_e\lesssim100$ pc) show no significant LIS line detections with stringent limit of $\mathrm{EW_\mathrm{LIS}}>-0.2$ \AA. This \red{result} is consistent with low column density (e.g., $\log(N_\mathrm{SiII})\lesssim13.5$) and/or low covering fraction ($C_f\lesssim0.1$), both of which are much lower than \red{previous measurements at $z\sim0-3$ and those for the less compact galaxies at $z\sim7-9$ in our sample}. We also identify that $\mathrm{EW_{LIS}}$ is correlated with H$\beta$-to-UV luminosity ratio and SFR surface density, suggesting that the weak LIS absorption in compact high-$z$ galaxies \red{may be driven by bursty and intense star formation or nuclear activity.}
    
    \item[2.] 
    For the five galaxies with multiple LIS lines detected, we simultaneously fit all the detected lines, accounting for the effects of both covering fraction and column densities, and derive abundance ratios. The [C/O] and [Si/O] ratios of our \red{sample} galaxies are comparable to those of DLA systems at $z\sim5-7$. Two galaxies with Fe \textsc{ii} line detections, GLASS-10021 and EGS-z7p2, show near-solar abundance ratios of $\mathrm{[Fe/Si]}\simeq-0.2$, suggesting that substantial Fe enrichment already occurred as early as $z\sim7$. To explore the origin of the Fe enrichment, we construct chemical evolution of models including CCSN + SN Ia\red{, CCSN + HN,} and CCSN + PISN yields. The observed near-solar ratios of [Al/Si] are inconsistent with the strong odd-even effect expected from PISNe, favoring SNe Ia. The SED fitting results of the two galaxies suggest young mass-weighted stellar age, possibly requiring prompt SNe Ia and dual starbursts.
\end{itemize}

%% Also note that the akcnowlodgment environment does not support long amounts of text. If you have a lot of people and institutions to acknowledge, do not use this command. Instead, create a new \section{Acknowledgments}.

\section*{Acknowledgements} 
\red{We are grateful to anonymous referee for the valuable comments that improved this manuscript. We thank Yuta Kageura, Makoto Ando, Yuichi Harikane, Chiaki Kobayashi, Xihan Ji, Matteo Messa, Hiroya Umeda, Eros Vanzella, and Hiroto Yanagisawa for the valuable discussions on this work. We thank Yuki Isobe for providing the line-spread functions of NIRSpec and for the useful discussions.}
This work is based on observations made with the NASA/ESA/CSA James Webb Space Telescope. The data were obtained from the Mikulski Archive for Space Telescopes at the Space Telescope Science Institute, which is operated by the Association of Universities for Research in Astronomy, Inc., under NASA contract NAS 5-03127 for JWST. \red{The calibration level 3 data in MAST are available at \dataset[10.17909/36bc-ej73]{https://doi.org/10.17909/36bc-ej73}. In our analysis, we begin with the level 2 versions.} These observations are associated with programs GO-9214 (SPURS), ERS-1345 (CEERS), GO-2561 (UNCOVER), GO-4111 (MegaScience), GTO-1180, 1181, 1210, 1264, 1286, 1287, 4540, GO-1895, 2514, 2674, 3577, 4762, 5398, and 6434 (JADES). We acknowledge the SPURS, CEERS, UNCOVER, MegaScience, and JADES teams led by Charlotte Mason \& Dan Stark, Steven L. Finkelstein, Ivo Labbé \& Rachel Bezanson, Katherine A. Suess, and Daniel Eisenstein \& Nora Lüetzgendorf, respectively, for developing their observation programs. Some of the data products presented herein were retrieved from the Dawn JWST Archive (DJA). DJA is an initiative of the Cosmic Dawn Center (DAWN), which is funded by the Danish National Research Foundation under grant DNRF140. We thank DJA for providing the reduced NIRSpec data. 
MN acknowledges the support by JSPS KAKENHI (25KJ0828) through Japan Society for the Promotion of Science (JSPS) and \red{by Japan Foundation for Promotion of Astronomy}. MO acknowledges the supports from the World Premier International Research Center Initiative (WPI Initiative), MEXT, Japan, the joint research program of the Institute for Cosmic Ray Research (ICRR), the University of Tokyo, and KAKENHI (21H04467, 25H00674) through JSPS. The English in this paper was partially refined with the assistance of ChatGPT (OpenAI).


%% citation or url to the code inside ()s when available

\software{NumPy \citep{Harris2020}, matplotlib \citep{Hunter2007}, SciPy \citep{Virtanen2020}, Astropy \citep{Astropy2013,Astropy2018,Astropy2022}, \texttt{grizli} \citep{Brammer2021,Brammer2023}, \texttt{msaexp} \citep{Brammer2023}, \texttt{Photutils} \citep{Bradley2025}, \texttt{Bagpipes} \citep{Carnall2018}, emcee \citep{Foreman2013}}.





%\clearpage
%\bibliographystyle{aasjournal}
%\bibliographystyle{aasjournalv7}
\bibliography{library.bib}



\clearpage
\appendix
\restartappendixnumbering
\section{Column Density and Covering Fraction Diagrams}
\label{Appendix:N-Cf}

\begin{figure*}[ht!]
    \centering
    \includegraphics[width=0.99\linewidth]{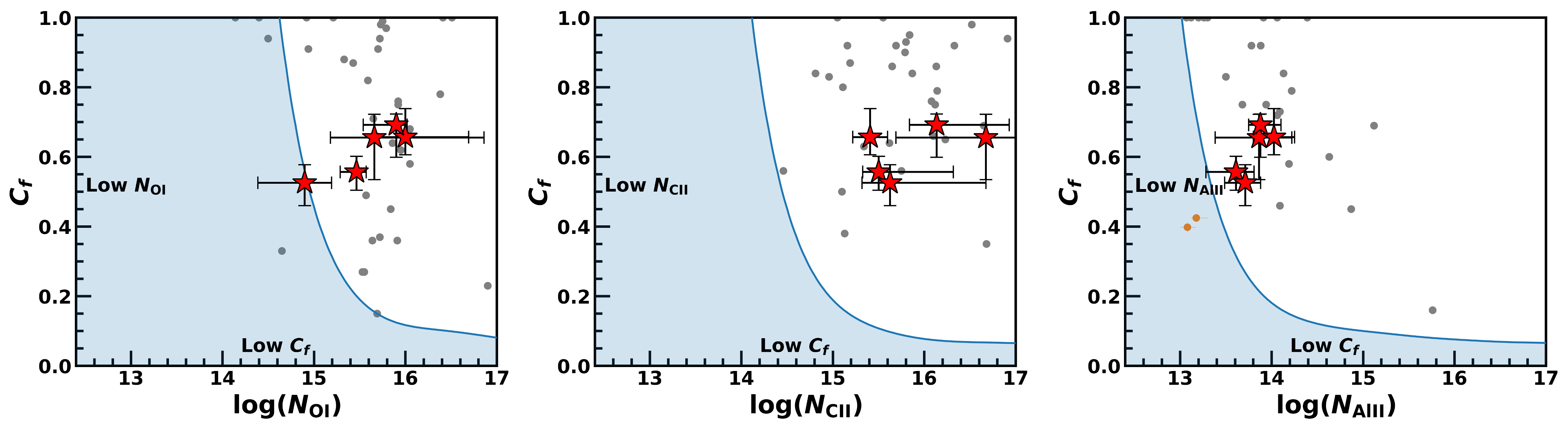}
    \caption{\red{Same as Figure \ref{fig:N-Cf}, but for O \textsc{i}, C \textsc{ii}, and Al \textsc{ii} column densities.}}
    \label{fig:N-Cf_appendix}
\end{figure*}

\red{In Section \ref{subsec:weak_LIS}, we discuss that the weak LIS absorption in compact galaxies at $z>7$ can be interpreted as low column density and/or covering fraction, as illustrated for Si \textsc{ii} in Figure \ref{fig:N-Cf}. In Figure \ref{fig:N-Cf_appendix}, we also present analogous $N$-$C_f$ diagrams for other ions of O \textsc{i}, C \textsc{ii}, and Al \textsc{ii}, for which suitable comparison measurements are available. Overall, the EW constraints for compact $z>7$ galaxies favor lower column densities and covering fractions than those of extended $z>7$ galaxies and lower-redshift galaxies across the different ions, supporting our conclusion.}

\end{document}

%% file: author_list.tex
\author[0009-0000-1999-5472]{Minami Nakane}
\affiliation{Institute for Cosmic Ray Research, The University of Tokyo, 5-1-5 Kashiwanoha, Kashiwa, Chiba 277-8582, Japan}
\affiliation{Department of Physics, Graduate School of Science, The University of Tokyo, 7-3-1 Hongo, Bunkyo, Tokyo 113-0033, Japan}
\email[show]{nakanem@icrr.u-tokyo.ac.jp}

\author[0000-0002-1049-6658]{Masami Ouchi}
\affiliation{National Astronomical Observatory of Japan, 2-21-1 Osawa, Mitaka, Tokyo 181-8588, Japan}
\affiliation{Institute for Cosmic Ray Research, The University of Tokyo, 5-1-5 Kashiwanoha, Kashiwa, Chiba 277-8582, Japan}
\affiliation{Department of Astronomical Science, SOKENDAI (The Graduate University for Advanced Studies), 2-21-1 Osawa, Mitaka, Tokyo, 181-8588, Japan}
\affiliation{Kavli Institute for the Physics and Mathematics of the Universe (WPI), The University of Tokyo, 5-1-5 Kashiwanoha, Kashiwa, Chiba 277-8583, Japan}
\email{ouchims@icrr.u-tokyo.ac.jp}

%% file: Table/table_sample.tex
\begin{table*}
    \caption{Sample in This Study.}
    \begin{tabular}{cccccccc}
    \hline
    \hline
    Name & R.A. & Decl. & $z_\mathrm{sys}$ & $M_\mathrm{UV}$ & $r_e$ & $\mu$ & Reference\\ 
    & & & & [mag] & [pc] & & \\
    (1) & (2) & (3) & (4) & (5) & (6) & (7) & (8)\\
    \hline
    GN-11 & 189.106030 & 62.242046 & 10.6034 & $-21.71_{-0.04}^{+0.02}$ & $67.7_{-5.8}^{+2.9}$ & - & Bu23, This work \\
    Gz9p3$^\dagger$ & 3.617169 & -30.425549 & 9.3111 & $-21.24_{-0.09}^{+0.06}$ & $409.2_{-31.6}^{+31.7}$ & 1.62 & Ch26, This work \\
    CEERS-1025 & 214.967547 & 52.932953 & 8.7163 & $-20.96_{-0.02}^{+0.02}$ & $53.4_{-4.6}^{+7.3}$ & - & This work \\
    CEERS-1019$^\dagger$ & 215.035391 & 52.890662 & 8.6778 & $-21.87_{-0.13}^{+0.08}$ & $102.6_{-8.9}^{+5.9\mathrm{a}}$ & - & This work \\
    GLASS-100003$^\dagger$ & 3.604509 & -30.380444& 7.8782 & $-20.58_{-0.13}^{+0.06}$ & $500.5_{-40.2}^{+46.0\mathrm{a}}$ & 1.34 & Na23, This work \\
    GLASS-10021$^\dagger$ & 3.608511 & -30.418541 & 7.2875 & $-20.91_{-0.06}^{+0.06}$ & $768.6_{-42.6}^{+7.8}$ & 1.72 & Na23, This work \\
    GN-z7p2 & 189.084150 & 62.222023 & 7.2021 & $-20.70_{-0.17}^{+0.15}$ & $64.7_{-14.2}^{+9.5\mathrm{a}}$ & - & This work \\
    EGS-z7p2 & 215.035614 & 52.892208 & 7.1934 & $-21.69_{-0.08}^{+0.05}$ & $846.0_{-32.2}^{+64.4}$ & - & This work \\
    \hline
    \end{tabular}
    \par
    \vspace{0.02\hsize}
    \footnotesize{(1): Name. (2): R.A. (3): Decl. (4): Systemic redshift. (5): Absolute UV magnitude corrected for magnification. (6): Half-light radius corrected for magnification. (7): Magnification factor. (8): References for systemic redshift and magnification factor (Bu23: \citealt{Bunker2023}, Ch26: \citealt{Chen2026}, Na23: \citealt{Nakajima2023}). $^\dagger$ UV absorption lines in these galaxies have also been analyzed in the literature \citep{Boyett2024,Chen2026,Pollock2026,VasanGC2026,Zhu2026}. $^\mathrm{a}$ The $r_e$ value of the brightest component is adopted (see Section \ref{subsubsec:morphology}).}
    \label{tab:sample}
\end{table*}

%% file: Table/table_line.tex
\begin{longrotatetable}
% \centering
\begin{table*}
% \begin{deluxetable*}
    \vspace*{-1cm}
    \caption{Line Measurements.}
    \hspace*{-3cm}
    \begin{tabular}{ccccccccc}
    \hline
    \hline
    Line & GN-z11 & Gz9p3 & CEERS-1025 & CEERS-1019 & GLASS-100003 & GLASS-10021 & GN-z7p2 & EGS-z7p2\\
    \hline
    \multicolumn{9}{c}{Absorption Line EW [\AA]} \\
    \hline
    Si \textsc{ii} $\lambda1260$ & $>-0.22$ & $-0.77_{-0.15}^{+0.05}$ & $>-0.31$ & $-0.73_{-0.16}^{+0.06}$ & $-2.04_{-0.40}^{+0.40}$ & $-1.04_{-0.32}^{+0.13}$ & $>-0.84$ & $-2.07_{-0.39}^{+0.24}$ \\
    O \textsc{i} $\lambda1302$ & $>-0.26$ & $-0.44_{-0.14}^{+0.04}$ & $>-0.27$ & $-0.26_{-0.10}^{+0.08}$ & $-1.23_{-0.26}^{+0.26}$ & $-1.32_{-0.30}^{+0.10}$ & $>-1.02$ & $-0.92_{-0.19}^{+0.19}$ \\
    Si \textsc{ii} $\lambda1304$ & $>-0.20$ & $-0.52_{-0.12}^{+0.08}$ & $>-0.19$ & $>-0.47$ & $>-1.08$ & $>-1.05$ & $>-0.77$ & $-1.13_{-0.25}^{+0.15}$ \\
    C \textsc{ii} $\lambda1334$ & $>-0.23$ & $-1.24_{-0.13}^{+0.10}$ & $>-0.24$ & $-1.13_{-0.14}^{+0.11}$ & $-1.77_{-0.42}^{+0.42}$ & $-1.38_{-0.35}^{+0.12}$ & $-0.89_{-0.27}^{+0.11}$ & $-2.75_{-0.26}^{+0.19}$ \\
    Si \textsc{iv} $\lambda1394$ & $-1.03_{-0.20}^{+0.12}$ & $-1.34_{-0.19}^{+0.06}$ & $>-0.49$ & $-1.05_{-0.08}^{+0.08}$ & $-0.75_{-0.19}^{+0.14}$ & - & $-0.88_{-0.27}^{+0.16}$ & $-1.72_{-0.34}^{+0.11}$ \\
    Si \textsc{iv} $\lambda1403$ & $-0.53_{-0.15}^{+0.18}$ & $-1.10_{-0.17}^{+0.06}$ & $>-0.29$ & $-1.00_{-0.12}^{+0.05}$ & $-0.81_{-0.18}^{+0.13}$ & - & $-1.00_{-0.28}^{+0.17}$ & $-1.56_{-0.22}^{+0.16}$ \\
    Si \textsc{ii} $\lambda1526$ & $>-0.46$ & $-0.56_{-0.15}^{+0.11}$ & $>-0.38$ & $-0.58_{-0.15}^{+0.09}$ & - & $-1.32_{-0.31}^{+0.04}$ & $>-0.99$ & $-1.66_{-0.17}^{+0.22}$ \\
    Fe \textsc{ii} $\lambda1608$ & $>-0.89$ & $>-0.63$ & $>-0.50$ & $>-0.57$ & $>-1.10$ & $-0.81_{-0.23}^{+0.17}$ & $>-0.94$ & $-0.95_{-0.32}^{+0.11}$ \\
    Al \textsc{ii} $\lambda1670$ & $>-0.38$ & $-0.50_{-0.20}^{+0.12}$ & $>-0.54$ & $-0.66_{-0.21}^{+0.08}$ & $-1.29_{-0.35}^{+0.26}$ & $-1.09_{-0.25}^{+0.19}$ & - & $-1.14_{-0.28}^{+0.17}$ \\
    Al \textsc{iii} $\lambda1854$ & $>-0.95$ & $-0.81_{-0.26}^{+0.21}$ & $>-0.70$ & $>-0.70$ & $>-1.14$ & $>-0.78$ & $>-1.20$ & $-1.61_{-0.44}^{+0.17}$ \\
    Al \textsc{iii} $\lambda1862$ & $>-1.61$ & $>-1.36$ & $>-0.69$ & $>-0.74$ & $>-1.18$ & $>-1.07$ & $>-1.19$ & $-1.13_{-0.39}^{+0.23}$ \\
    Fe \textsc{ii} $\lambda2600$ & $>-2.19$ & $>-1.87$ & $>-2.82$ & $>-2.24$ & $>-4.96$ & $-2.92_{-0.83}^{+0.83}$ & $>-2.05$ & $-3.15_{-0.80}^{+0.48}$ \\
    Mg \textsc{ii} $\lambda\lambda2796,2803^\mathrm{a}$ & - & - & - & - & - & - & - & $-5.50_{-1.86}^{+0.62}$ \\
    LIS$^\mathrm{b}$ & $>-0.21$ & $-0.87_{-0.10}^{+0.05}$ & $>-0.19$ & $-0.81_{-0.11}^{+0.04}$ & $-1.86_{-0.33}^{+0.25}$ & $-1.26_{-0.22}^{+0.04}$ & $-0.49_{-0.12}^{+0.08}$ & $-2.20_{-0.15}^{+0.15}$ \\
    \hline
    \multicolumn{9}{c}{Emission Line Flux [$10^{-19}$ erg s$^{-1}$ cm$^{-2}$ \AA$^{-1}$]} \\
    \hline
    
    Ly$\alpha$ & $16.53_{-0.84}^{+0.63}$ & $<0.96$ & $7.15_{-0.75}^{+0.45}$ & $23.96_{-2.08}^{+2.08\mathrm{c}}/28.26_{-3.92}^{+1.57\mathrm{d}}$ & $<2.19$ & $12.39_{-5.23}^{+2.09}$ & $15.95_{-2.28}^{+1.37}$ & $<4.35$ \\
    N \textsc{iv}] $\lambda1483$ & $<2.16$ & $<0.90$ & $<0.70$ & $5.72_{-0.49}^{+0.16}$ & - & - & $<1.73$ & $<2.45$ \\
    N \textsc{iv}] $\lambda1486$ & $9.19_{-0.52}^{+0.39}$ & $<1.00$ & $<1.17$ & $25.93_{-0.66}^{+0.22}$ & - & - & $<1.58$ & $<2.98$ \\
    He \textsc{ii} $\lambda1640$ & $<4.18$ & $<0.96$ & $2.78_{-0.68}^{+0.41}$ & $5.99_{-0.60}^{+0.36}$ & $5.24_{-0.83}^{+0.50}$ & $3.47_{-1.18}^{+0.94}$ & - & $<2.66$ \\
    O \textsc{iii}] $\lambda1661$ & $<1.95$ & $<1.04$ & $<1.35$ & $2.99_{-0.37}^{+0.27}$ & $2.64_{-0.57}^{+0.43}$ & $<7.24$ & - & $<1.76$ \\
    O \textsc{iii}] $\lambda1666$ & $2.30_{-0.47}^{+0.47}$ & $<1.37$ & $<1.68$ & $7.05_{-0.46}^{+0.35}$ & $7.75_{-0.97}^{+0.58}$ & $5.54_{-1.70}^{+0.68}$ & - & $<2.80$ \\
    Si \textsc{iii}] $\lambda1883$ & $<2.47$ & $<1.08$ & $<1.41$ & $1.98_{-0.33}^{+0.44}$ & $2.38_{-1.12}^{+0.37}$ & $<6.02$ & $<1.18$ & $<2.11$ \\
    Si \textsc{iii}] $\lambda1892$ & $<1.94$ & $<1.21$ & $<1.10$ & $1.87_{-0.42}^{+0.42}$ & $1.56_{-0.92}^{+0.55}$ & $<5.71$ & $<1.26$ & $<2.76$ \\
    C \textsc{iii}] $\lambda1907$ & $3.94_{-0.65}^{+0.22}$ & $1.52_{-0.38}^{+0.23}$ & $<2.12$ & $9.86_{-0.65}^{+0.43}$ & $8.21_{-0.67}^{+0.67}$ & $11.36_{-1.36}^{+1.36}$ & $1.42_{-0.38}^{+0.38}$ & $<2.93$ \\
    C \textsc{iii}] $\lambda1909$ & $3.29_{-0.57}^{+0.28}$ & $<1.67$ & $0.91_{-0.35}^{+0.26}$ & $9.59_{-0.79}^{+0.40}$ & $6.15_{-0.84}^{+0.42}$ & $11.89_{-1.97}^{+0.79}$ & $1.06_{-0.38}^{+0.30}$ & $<2.36$ \\
    Mg \textsc{ii} $\lambda2796^\mathrm{a}$ & - & - & - & $3.13_{-0.65}^{+0.39}$ & - & - & - & - \\
    Mg \textsc{ii} $\lambda2803^\mathrm{a}$ & - & - & - & $1.84_{-0.79}^{+0.26}$ & - & - & - & - \\
    {}[O \textsc{ii}] $\lambda3727$ & $4.59_{-0.78}^{+0.31}$ & $2.25_{-0.52}^{+0.31}$ & $2.95_{-0.57}^{+0.34}$ & $11.90_{-0.62}^{+0.62}$ & $2.40_{-0.95}^{+0.32}$ & $8.39_{-2.24}^{+0.75}$ & $5.30_{-1.87}^{+0.93}$ & $12.21_{-3.16}^{+0.45}$ \\
    {}[O \textsc{ii}] $\lambda3729$ & $<3.09$ & $2.57_{-0.65}^{+0.11}$ & $<2.48$ & $6.51_{-0.60}^{+0.60}$ & $3.49_{-0.77}^{+0.46}$ & $12.41_{-2.20}^{+0.88}$ & $<7.84$ & $11.73_{-1.90}^{+1.52}$ \\
    {}[Ne \textsc{iii}] $\lambda3869$ & $8.63_{-1.14}^{+0.38}$ & $2.67_{-0.74}^{+0.44}$ & $2.61_{-0.63}^{+0.63}$ & $36.88_{-1.09}^{+0.36}$ & $9.04_{-0.90}^{+0.67}$ & $16.78_{-2.53}^{+1.52}$ & $5.95_{-1.16}^{+1.16}$ & $6.15_{-2.23}^{+1.34}$ \\
    He \textsc{i}+H8 & $2.50_{-0.83}^{+0.28}$ & $<1.24$ & $<1.57$ & $8.58_{-0.79}^{+0.48}$ & $2.88_{-0.79}^{+0.48}$ & $<5.87$ & $<3.67$ & $<3.40$ \\
    {}[Ne \textsc{iii}] $\lambda3967$+H7 & $7.35_{-1.29}^{+0.78}$ & $<2.33$ & $<2.25$ & $22.91_{-1.10}^{+0.44}$ & - & $<8.28$ & $<5.09$ & $<4.57$ \\
    H$\delta$ & $4.91_{-1.39}^{+0.46}$ & $1.69_{-0.27}^{+0.55}$ & $1.56_{-0.58}^{+0.19}$ & $17.32_{-1.06}^{+0.42}$ & - & $7.31_{-1.74}^{+0.29}$ & $6.23_{-1.54}^{+1.16}$ & $<4.43$ \\
    H$\gamma$ & $7.26_{-2.80}^{+1.68}$ & $1.77_{-0.42}^{+0.42}$ & $3.92_{-0.76}^{+0.57}$ & $25.09_{-0.99}^{+0.40}$ & $6.33_{-0.64}^{+0.38}$ & $15.93_{-2.19}^{+0.36}$ & $2.90_{-0.92}^{+0.55}$ & $5.05_{-1.31}^{+0.98}$ \\
    {}[O \textsc{iii}] $\lambda4363$ & $<6.95$ & $<1.96$ & $<2.48$ & $9.66_{-0.67}^{+0.50}$ & $2.69_{-0.71}^{+0.42}$ & $5.66_{-1.57}^{+0.63}$ & $<3.51$ & $<3.96$ \\
    H$\beta$ & - & $4.49_{-1.05}^{+0.63}$ & $7.62_{-0.65}^{+0.87}$ & $45.04_{-2.06}^{+1.24\mathrm{c}}/13.81_{-2.26}^{+1.69\mathrm{d}}$ & $13.01_{-1.01}^{+0.61}$ & $35.40_{-1.96}^{+0.65}$ & $6.51_{-1.03}^{+0.62}$ & $12.13_{-1.36}^{+1.02}$ \\
    {}[O \textsc{iii}] $\lambda4959$ & - & $12.44_{-0.41}^{+0.19}$ & $11.20_{-0.46}^{+0.22}$ & $123.21_{-3.94}^{+1.45\mathrm{c}}/35.72_{-3.65}^{+1.54\mathrm{d}}$ & $34.17_{-0.75}^{+0.41}$ & $79.86_{-0.52}^{+0.67}$ & $18.98_{-0.49}^{+0.24}$ & $17.31_{-0.67}^{+0.27}$ \\
    {}[O \textsc{iii}] $\lambda5007$ & - & $37.08_{-1.23}^{+0.57}$ & $33.39_{-1.37}^{+0.65}$ & $367.16_{-11.74}^{+4.31\mathrm{c}}/106.45_{-10.87}^{+4.60\mathrm{d}}$ & $101.83_{-2.23}^{+1.22}$ & $237.98_{-1.54}^{+1.98}$ & $56.57_{-1.45}^{+0.71}$ & $51.60_{-2.00}^{+0.80}$ \\
    \hline
    \end{tabular}
    \par
    \vspace{0.02\hsize}
    \footnotesize{$^\mathrm{a}$ Mg \textsc{ii} line fluxes/EWs are measured only for detections since these lines can be observed either as emission or absorption lines. $^\mathrm{b}$ Mean equivalent widths of Si \textsc{ii} $\lambda\lambda1260,1526$ and C \textsc{ii} $\lambda1334$ lines. $^\mathrm{c}$ Narrow component fluxes. $\mathrm{d}$ Broad component fluxes.}
    \label{tab:line}
\end{table*}
% \end{deluxetable*}
\end{longrotatetable}

%% file: Table/table_absorption.tex
\begin{table*}
    \caption{Covering Fraction and Column Densities of LIS Absorption Lines.}
    \begin{tabular}{cccccccc}
    \hline
    \hline
    Name & $C_f$ & $\log(N_\mathrm{OI})$ & $\log(N_\mathrm{CII})$ & $\log(N_\mathrm{SiII})$ & $\log(N_\mathrm{AlI})$ & $\log(N_\mathrm{FeII})$ & $\log(N_\mathrm{MgII})$ \\ 
    (1) & (2) & (3) & (4) & (5) & (6) & (7) & (8)\\
    \hline
    Gz9p3 & $0.56_{-0.05}^{+0.05}$ & $15.47_{-0.18}^{+0.11}$ & $15.50_{-0.17}^{+0.82}$ & $14.94_{-0.14}^{+0.14}$ & $13.61_{-0.33}^{+0.20}$ & - & - \\
    CEERS-1019 & $0.53_{-0.07}^{+0.05}$ & $14.90_{-0.51}^{+0.29}$ & $15.63_{-0.31}^{+1.05}$ & $14.65_{-0.18}^{+0.12}$ & $13.71_{-0.22}^{+0.17}$ & - & - \\
    GLASS-100003 & $0.66_{-0.12}^{+0.07}$ & $15.66_{-0.48}^{+1.20}$ & $16.67_{-0.98}^{+0.61}$ & $15.06_{-0.47}^{+0.47}$ & $13.86_{-0.48}^{+0.36}$ & - & - \\
    GLASS-10021 & $0.66_{-0.05}^{+0.08}$ & $16.00_{-0.32}^{+0.69}$ & $15.41_{-0.19}^{+0.19}$ & $15.14_{-0.15}^{+0.18}$ & $14.03_{-0.23}^{+0.23}$ & $15.01_{-0.14}^{+0.21}$ & - \\
    EGS-z7p2 & $0.69_{-0.09}^{+0.03}$ & $15.90_{-0.36}^{+0.12}$ & $16.14_{-0.30}^{+0.79}$ & $15.28_{-0.19}^{+0.51}$ & $13.88_{-0.13}^{+0.23}$ & $15.06_{-0.16}^{+0.16}$ & $14.63_{-0.34}^{+1.19}$ \\
    
    \hline
    \end{tabular}
    \par
    \vspace{0.02\hsize}
    \footnotesize{(1): Name. (2): Covering fraction. (3): O \textsc{i} column density. (4): C \textsc{ii} column density. (5): Si \textsc{ii} column density. (6): Al \textsc{ii} column density. (7): Fe \textsc{ii} column density. (8): Mg \textsc{ii} column density.}
    \label{tab:absorption}
\end{table*}

%% file: Table/table_nebular.tex
\begin{table*}
    \centering
    \caption{\red{Nebular Properties.}}
    \begin{tabular}{cccccccccc}
    \hline
    \hline
    Name & $12+\log(\mathrm{O/H})$ & $E(B-V)$ & $\log(n_e)$ & $T_e$([O\textsc{iii}]) & $T_e$([O\textsc{ii}]) & \red{$\log{\mathrm{R3}}$} & \red{$\log{\mathrm{R2}}$} & $\beta$ & Method\\ 
    & & & [$\mathrm{cm^{-2}}$] & [$10^4$ K] & [$10^4$ K] & & & & \\
    (1) & (2) & (3) & (4) & (5) & (6) & (7) & (8) & (9) & (10) \\
    \hline
    %GN-z11 & \\
    Gz9p3 & $7.95_{-0.22}^{+0.28}$ & $0.05_{-0.03}^{+0.27}$ & $2.58_{-1.56}^{+0.48}$ & - & - & $0.87_{-0.06}^{+0.02}$ & $0.05_{-0.09}^{+0.09}$ & $-1.95_{-0.02}^{+0.01}$ & Strong line \\
    CEERS-1025 & $7.81_{-0.29}^{+0.26}$ & $0.28_{-0.26}^{+0.19}$ & $4.93_{-1.24}^{+0.45}$ & - & - & $0.81_{-0.05}^{+0.03}$ & $-0.07_{-0.13}^{+0.09}$ & $-2.40_{-0.03}^{+0.02}$ & Strong line \\
    CEERS-1019 & $7.84_{-0.05}^{+0.03}$ & $0.00_{-0.00}^{+0.02}$ & $3.63_{-0.28}^{+0.16}$ & $1.68_{-0.06}^{+0.05}$ & $1.37_{-0.26}^{+0.26}$ & $0.91_{-0.02}^{+0.01\mathrm{a}}$ & $-0.37_{-0.03}^{+0.02\mathrm{a}}$ & $-1.55_{-0.01}^{+0.01}$ & Direct $T_e$ \\
    GLASS-100003 & $7.76_{-0.14}^{+0.07}$ & $0.03_{-0.02}^{+0.21}$ & $0.48_{-0.43}^{+1.47}$ & $1.81_{-0.21}^{+0.15}$ & $1.56_{-0.52}^{+0.22}$ & $0.87_{-0.01}^{+0.04\mathrm{a}}$ & $-0.32_{-0.04}^{+0.16\mathrm{a}}$ & $-2.23_{-0.03}^{+0.02}$ & Direct $T_e$ \\
    GLASS-10021 & $7.71_{-0.12}^{+0.09}$ & $0.45_{-0.13}^{+0.11}$ & $0.04_{-0.00}^{+1.68}$ & $1.83_{-0.19}^{+0.15}$ & $1.63_{-0.50}^{+0.21}$ & $0.80_{-0.02}^{+0.01\mathrm{a}}$ & $0.02_{-0.15}^{+0.02\mathrm{a}}$ & $-2.43_{-0.02}^{+0.02}$ & Direct $T_e$ \\
    GN-z7p2 & $7.98_{-0.19}^{+0.25}$ & $0.01_{-0.00}^{+0.30}$ & $3.78_{-0.72}^{+2.04}$ & - & - & $0.88_{-0.05}^{+0.03}$ & $0.07_{-0.10}^{+0.13}$ & $-2.45_{-0.03}^{+0.03}$ & Strong line \\
    EGS-z7p2 & $8.74_{-0.17}^{+0.02}$ & $0.09_{-0.08}^{+0.26}$ & $2.70_{-1.32}^{+0.48}$ & - & - & $0.62_{-0.04}^{+0.03}$ & $0.36_{-0.06}^{+0.13}$ & $-1.51_{-0.02}^{+0.01}$ & Strong line \\
    
    \hline
    \end{tabular}
    \par
    \vspace{0.02\hsize}
    \footnotesize{(1): Name. (2): Gas-phase metallicity. (3): Color excess. (4): Electron density based on [O \textsc{ii}] $\lambda\lambda3727,3729$. (5): Electron temperature of [O \textsc{iii}]. (6): Electron temperature of [O \textsc{ii}]. (7): [O \textsc{iii}] $\lambda\lambda4959,5007$/H$\beta$ flux ratio. (8): [O \textsc{ii}] $\lambda\lambda3727,3729$/H$\beta$ flux ratio. (9) Method for connecting metallicity and flux ratios. \red{$^\mathrm{a}$ The R3 and R2 ratios are derived from the fluxes presented in Table \ref{tab:line}.}}
    \label{tab:nebular}
\end{table*}

%% file: Table/table_SED.tex
\begin{table*}
    \caption{Stellar Populations.}
    \begin{tabular}{ccccccc}
    \hline
    \hline
    Name & $\log M_*$ & Age & SFR & $A_\mathrm{v}$ & $Z$ & $\log U$ \\ 
    & [$M_\odot$] & [Myr] & [$M_\odot\ \mathrm{yr^{-1}}$] &[mag] & [$Z_\odot$] &\\
    (1) & (2) & (3) & (4) & (5) & (6) & (7) \\
    \hline
    GN-z11 & $8.73_{-0.04}^{+0.04}$ & $41.70_{-4.92}^{+24.77}$ & $11.10_{-1.33}^{+1.15}$ & $0.05_{-0.01}^{+0.02}$ & $0.17_{-0.03}^{+0.03}$ & $-1.52_{-0.27}^{+0.37}$ \\
    Gz9p3 & $9.23_{-0.02}^{+0.02}$ & $20.02_{-0.42}^{+2.28}$ & $20.34_{-0.86}^{+0.88}$ & $0.48_{-0.08}^{+0.09}$ & $0.16_{-0.02}^{+0.02}$ & $-0.83_{-0.19}^{+0.19}$ \\
    CEERS-1025 & $8.65_{-0.06}^{+0.04}$ & $20.97_{-2.05}^{+14.57}$ & $5.24_{-0.72}^{+0.55}$ & $0.66_{-0.12}^{+0.09}$ & $0.06_{-0.04}^{+0.05}$ & $-0.65_{-0.18}^{+0.11}$ \\
    CEERS-1019 & $10.32_{-0.02}^{+0.01}$ & $316.42_{-6.98}^{+5.08}$ & $34.26_{-3.92}^{+5.11}$ & $0.21_{-0.01}^{+0.01}$ & $0.58_{-0.03}^{+0.03}$ & $-0.51_{-0.01}^{+0.00}$ \\
    GLASS-100003 & $9.07_{-0.01}^{+0.01}$ & $38.80_{-0.55}^{+3.87}$ & $14.60_{-0.38}^{+0.33}$ & $0.20_{-0.00}^{+0.00}$ & $0.24_{-0.01}^{+0.01}$ & $-0.52_{-0.03}^{+0.02}$ \\
    GLASS-10021 & $8.86_{-0.01}^{+0.01}$ & $19.46_{-0.41}^{+2.64}$ & $8.49_{-0.22}^{+0.20}$ & $0.25_{-0.03}^{+0.03}$ & $0.23_{-0.01}^{+0.01}$ & $-0.80_{-0.16}^{+0.16}$ \\
    GN-z7p2 & $8.89_{-0.05}^{+0.04}$ & $370.86_{-35.47}^{+21.55}$ & $1.50_{-0.57}^{+0.77}$ & $0.40_{-0.03}^{+0.03}$ & $0.11_{-0.06}^{+0.07}$ & $-0.62_{-0.18}^{+0.09}$ \\
    EGS-z7p2 & $9.48_{-0.04}^{+0.03}$ & $13.26_{-1.89}^{+4.38}$ & $34.82_{-3.06}^{+2.82}$ & $0.95_{-0.04}^{+0.03}$ & $0.75_{-0.02}^{+0.02}$ & $-1.91_{-0.18}^{+0.23}$ \\

    \hline
    \end{tabular}
    \par
    \vspace{0.02\hsize}
    \footnotesize{(1): Name. (2): Stellar mass corrected for magnification. (3): Mass-weighted stellar age. (4): Star formation rate corrected for magnification. (5): V-band attenuation. (6): Stellar metallicity. (7): Ionization parameter. }
    \label{tab:SED}
\end{table*}

%% file: Table/table_abundance.tex
\begin{table*}
    % \centering
    \caption{Abundance Ratio Measurements.}
    \begin{tabular}{ccccccc}
    \hline
    \hline
    Name & [C/O] & [Si/O] & [Fe/Si] & [Al/Si] & [O/Fe] & [Mg/Fe]\\ 
    (1) & (2) & (3) & (4) & (5) & (6) & (7)\\
    \hline
    Gz9p3 & $0.35_{-0.38}^{+0.75}$ & $0.62_{-0.21}^{+0.21}$ & $<0.12$ & $-0.34_{-0.37}^{+0.29}$ & $>-0.78$ & - \\
    CEERS-1019 & $1.11_{-0.37}^{+1.48}$ & $0.89_{-0.43}^{+0.52}$ & $<0.41$ & $0.19_{-0.22}^{+0.28}$ & $>-1.34$ & - \\
    GLASS-100003 & $0.58_{-0.94}^{+1.32}$ & $0.43_{-1.63}^{+0.72}$ & $<0.00$ & $-0.35_{-0.97}^{+0.65}$ & $>-0.58$ & - \\
    GLASS-10021 & $-0.35_{-0.90}^{+0.30}$ & $0.30_{-0.76}^{+0.35}$ & $-0.15_{-0.23}^{+0.28}$ & $0.06_{-0.47}^{+0.13}$ & $-0.24_{-0.42}^{+0.71}$ & - \\
    EGS-z7p2 & $0.56_{-0.47}^{+0.93}$ & $0.69_{-0.35}^{+0.69}$ & $-0.23_{-0.67}^{+0.19}$ & $-0.38_{-0.61}^{+0.26}$ & $-0.49_{-0.32}^{+0.32}$ & $-0.44_{-0.47}^{+1.17}$ \\        
    \hline
    \end{tabular}
    \par
    \vspace{0.02\hsize}
    \footnotesize{(1): Name. (2) [C/O] ratio. (3) [Si/O] ratio. (4) [Fe/Si] ratio and $3\sigma$ upper limit. (5) [Al/Si] ratio. (6) [O/Fe] ratio and $3\sigma$ lower limit. (7) [Mg/Fe] ratio.}
    \label{tab:abundance}
\end{table*}

%Mg/Fe